# A Community Science Case for E-ELT HIRES

R. Maiolino[1,2], M. Haehnelt[2,3], M.T. Murphy[4], D. Queloz[1], L. Origlia[5], J. Alcala[6], Y. Alibert[7], P.J. Amado[8], C. Allende Prieto[9], M. Ammler-von Eiff[10], M. Asplund[11], M. Barstow[12], G. Becker[2,3], X. Bonfils[13], F. Bouchy[14], A. Bragaglia[5], M.R. Burleigh[12], A. Chiavassa[15], D. A. Cimatti[16], M. Cirasuolo[17], S. Cristiani[18], V. D'Odorico[18], D. Dravins[19], E. Emsellem[20], J. Farihi[3], P. Figueira[21], J. Fynbo[22], B.T. Gänsicke[23], M. Gillon[24], B. Gustafsson[25], V. Hill[26], G. Israelyan[9], A. Korn[25], S. Larsen[27], P. De Laverny[28], J. Liske[20], C. Lovis[29], A. Marconi[30], C. Martins[21], P. Molaro[18], B. Nisini[31], E. Oliva[32], P. Petitjean[33], M. Pettini[3], A. Recio Blanco[26], R. Rebolo[9], A. Reiners[34], C. Rodriguez-Lopez[35], N. Ryde[25], N.C. Santos[21], S. Savaglio[36], I. Snellen[37], K. Strassmeier[38], N. Tanvir[39], L. Testi[20], E. Tolstoy[40], A. Triaud[29], L. Vanzi[41], M. Viel[18], M. Volonteri[33]

*(Affliations can be found after the references)*

## Abstract

Building on the experience of the high-resolution community with the suite of VLT high-resolution spectrographs, which has been tremendously successful, we outline here the (science) case for a high-fidelity, high-resolution spectrograph with wide wavelength coverage at the E-ELT.

Flagship science drivers include:
- the study of exo-planetary atmospheres with the prospect of the detection of signatures of life on rocky planets,
- the chemical composition of planetary debris on the surface of white dwarfs,
- the spectroscopic study of protoplanetary and proto-stellar disks,
- the extension of Galactic archaeology to the Local Group and beyond,
- spectroscopic studies of the evolution of galaxies with samples that, unlike now, are no longer restricted to strongly star-forming and/or very massive galaxies,
- the unraveling of the complex roles of stellar and AGN feedback for the supply and retention of the baryonic component of galaxies across the full range of galaxy masses, morphologies and a wide range of redshift, with the help of IGM tomography at high spatial resolution,
- the study of the chemical signatures imprinted by population III stars on the IGM during the epoch of reionization,
- the exciting possibility of paradigm-changing contributions to fundamental physics due to the precision afforded by Laser Frequency Comb (LFC) calibrated high-fidelity spectroscopy.

The requirements of these science cases can be met by a stable instrument with a spectral resolution of R~100,000 and broad, simultaneous spectral coverage extending from 0.37μm to 2.5μm. Most science cases do not require spatially resolved information, and can be pursued in seeing-limited mode, although some of them would benefit by the E-ELT diffraction limited resolution. Some multiplexing would also be beneficial for some of the science cases.

HIRES will ensure the continued competiveness of the European high resolution community in the E-ELT era and in this way will largely enhance the overall competiveness of the E-ELT.

# Executive Summary

This White Paper summarizes the science cases and the associated instrument requirements for a high resolution spectrograph (HIRES) on the E-ELT.
The emerging favored instrument concept is that of a versatile instrument, capable of pursuing a number of key, outstanding science cases.

The characterization of exoplanets is one of the outstanding key science cases for HIRES. The focus will be on characterizing exo-planet atmospheres over a wide range of masses, from Neptune-like down to Earth-like including those in the habitable zones, in terms of chemical composition, stratification and weather. The ultimate goal is the detection of signatures of life. The extremely high signal-to-noise required to detect the exo-planet atmospheric signatures has paradoxically pushed this area into the "photon-starved" regime with current facilities, making the collecting area of the E-ELT essential for achieving the ambitious goals. The key requirements for this science case are a spectral resolution R~100,000 (primarily to disentangle the exoplanet atmospheric features from the telluric absorption lines of our atmosphere and to increase detection capability by allowing the detection of narrow lines, but also to trace different layers of the atmosphere and exoplanet weather), a wide wavelength range (0.37-2.5 μm), high stability of the PSF on the detector during planetary transits and high flat-fielding accuracy. A polarimetric mode would further enhance the exoplanet diagnostic capabilities of HIRES, especially for the detection of bio-signatures. Similar capabilities (especially in the blue part of the spectrum) are required for enabling the detection of planetary debris on the surface of white dwarfs, which is an alternative exciting and rapidly growing technique to trace the composition of exoplanets. In the ELT era, radial velocity studies will not only focus on the detection of exo-Earths, but also on the detection of weak and rare time-limited signals, like the Rossiter-McLaughlin effect. The latter will require a stability of the spectrometer of 10 cm $s^{-1}$.

HIRES, as discussed here, has also the unique capability of revealing the dynamics, chemistry and physical conditions of the inner regions of the accretion disks and protoplanetary disks of young stellar objects, hence providing unprecedented constraints on the physics of star formation, jet launching mechanisms and planet formation. To achieve these goals we propose here that in the near-IR the instrument's high resolution (R~100,000) is exploited with spatially resolved information (possibly with an IFU mode) at the diffraction limit of the E-ELT. This science case would also benefit from a polarimetric mode, which would provide information on the magnetic field in the inner regions of the accretion disk.

On the front of stellar spectroscopy HIRES will deliver for the very first time the high-resolution and high-quality spectra (S/N>100) required to trace in detail the chemical enrichment pattern of solar-type and cooler dwarf stars out to distances of several kpc, thus sampling most of the Galactic disk and bulge, or sub-giants and red giants in the outer Galactic Halo and in neighbouring dwarf galaxies. With a spectral resolution R~100,000 and a broad spectral coverage (0.37-2.5 μm), the detailed chemical mapping of multiple elements and isotopes, through HIRES spectra, will reveal the origin and the formation history of ancient stars belonging to different Galactic components. This will be crucial for the extremely low metallicity stars, whose photospheres may trace the chemical abundances resulting from the enrichment of the first population of stars (PopIII).
HIRES will also be an extremely efficient machine to trace the metal enrichment pattern and dynamics of extragalactic star clusters and resolved stellar populations, hence tracing the star formation history in other galaxies, if enabled with some multiplexing capability (~5-10 objects over a FoV of a few arcmin) with intermediate spectral resolution (R~20,000) sampling the full spectral range from 0.37 μm to 2.5 μm.

In the context of galaxy formation and cosmology, one of the most exciting prospects for HIRES is the detection of elements synthesized by the first stars in the Universe. HIRES will probably be the first facility that will detect the fingerprint of PopIII stars by measuring the chemical enrichment typical of this population in the Inter-Galactic (IGM) and Inter-Stellar Medium (ISM) in the foreground of Quasars, GRBs and Super-Luminous Supernovae at high redshift, probing in this way the epoch of reionization. These observations will reveal the nature and physical properties of the first stars that populated the Universe. The high spectral resolution of HIRES will also allow astronomers to trace in detail the history of the reionization process of the Universe and the subsequent thermal history of the IGM. To reach these exciting science goals with HIRES requires a spectral resolution R>50,000 and a spectral coverage extending from about 6000Å to 2.5 μm.

If enabled with some multiplexing capability (5-10 objects), HIRES will also be able to obtain a three-dimensional map of the cosmic web of the IGM at high redshift, by probing absorption systems towards multiple lines of sight on scales of a few arcminutes. Most importantly, if the simultaneous wavelength coverage extends from 4000Å to 2.5μm, HIRES will have the unique and exceptional capability of obtaining a three-dimensional map of the distribution of metals in the IGM, which would be a unique probe of the enrichment process of the Universe.

The same capabilities (i.e. intermediate spectral resolution, wide simultaneous spectral coverage and multiplexing of 5-10 over a field of view of a few arcminutes) are also required to investigate the processes driving the evolution of massive early type galaxies, during the epochs of their formation and assembly (z~1-3), which is still a major unsolved puzzle beyond reach of current facilities.

If equipped with an IFU sampling the ELT diffraction limit, HIRES, with its high spectral resolution (R~100,000) will be the only tool to measure the low mass end of supermassive black hole in galactic nuclei, which bears signature of primordial black hole seeds.

Perhaps most exciting, HIRES will be an instrument capable of addressing issues that go beyond the limited field of Astronomy, breaking into the domain of "fundamental physics". In particular, HIRES will provide the most accurate constraints on a possible variation of the fundamental constants of nature, and in particular of the fine structure constant $\alpha$ and of the proton-to-electron mass ratio $\mu$, some of the most exciting topics in physics. The latter measurement in particular will be enabled by a spectral resolution R~100,000, and high efficiency in the blue part of the spectrum.

HIRES will also deliver the most accurate measurement of the deuterium abundance that, compared with the value obtained from the CMB measurements (from Planck), will provide stringent constraints on models of non-standard physics. By measuring the redshift drift-rate dz/dt of absorption features in distant QSOs, HIRES will be the only instrument capable of obtaining a direct, non-geometric, completely model-independent measurement of the Universe's expansion history (the "Sandage test"). This should be regarded as (the beginning of) a legacy experiment, lasting several decades. However, in addition to high-spectral resolution (R~100,000), this measurement requires a very accurate *absolute* wavelength calibration of about 2 cm $s^{-1}$ (which can be achieved with laser comb technology and high fiber scrambling efficiency) as well as excellent stability, of the order of 2 cm $s^{-1}$, over the duration of an observing night.

In summary, the various science cases result in the following set of requirements: a primary high-resolution observing mode with R~100,000 and a simultaneous wavelength range from 0.37 mm to 2.5 μm (although the extension to 0.33 mm is desirable for some cases). For most science cases a stability of about 10 cm $s^{-1}$ and an accuracy of the relative wavelength calibration of 1 m $s^{-1}$ are sufficient. The exo-planet radial velocity cases also require a wavelength accuracy down to 10 cm $s^{-1}$. The Sandage test requires a stability as good as 2 cm $s^{-1}$ over the duration of a night, while the latter also requires an *absolute* wavelength calibration of 2 cm $s^{-1}$.
The science cases of mapping the metals in the IGM, galaxy evolution and extragalactic star clusters would greatly benefit from having, with the same wide spectral coverage (0.37-2.5 μm), a moderate multiplexing capability (5-10 objects within a FoV of a few arcminutes) with a moderate spectral resolution mode (R~10,000-50,000). Most of the extragalactic, high-z science cases require an accurate subtraction of the sky background, to better than 1%.

We suggest that these various requirements can be met well with a highly modular, low-risk instrument concept, where different independent modules enable different observing modes and give access to different wavelength ranges.

# 1. Introduction

High resolution spectroscopy has been, during the past twenty years, a rapidly expanding area enabling major, fundamental progresses in most fields of astrophysics, but also in more general areas of fundamental physics. European astronomers have been leading several of these fields as well as many of the major discoveries thanks to the four high-resolution spectrographs at the VLT (UVES, CRIRES, FLAMES, X-shooter, covering the full wavelength range from the near-UV at the atmospheric cut-off to 2.5μm) and HARPS at the 3.6m telescope. This suite is soon to be joined by the LFC-calibrated high-fidelity spectrograph ESPRESSO. This impressive high-resolution capability of the ESO telescopes has delivered excellent science in a wide range of fields from the discovery of exoplanets and the characterization of their atmospheres, stellar abundances, star and planet formation, Galactic Archaeology in our own galaxy, the study of supernovae and GRBs, to studies of galaxy evolution and the physical state of the IGM, and to interesting constraints on the nature of dark matter and the variation of fundamental constants, to name just a few. More than 40% of the scientific output of the VLT is based on its suite of high-resolution spectrographs (Grothkopf & Meakins 2013). The scientific output of these high-resolution spectrographs rises well above 50% when including the La Silla telescopes (HARPS).

It is clear that in the area of high resolution spectroscopy, where "photon starving" is the main limiting factor, the discovery space enabled by larger telescopes will be huge. Indeed, ESO had commissioned nine phase A studies for E-ELT instrument concepts that included one optical (CODEX) and one near-IR high-resolution spectrograph (SIMPLE). Based on the result of these phase A studies, ESO published in 2011 an instrumentation roadmap that foresees a high-resolution spectrograph, E-ELT HIRES, as either instrument Nr 4 or Nr 5.

In the meantime the CODEX and SIMPLE consortia have merged and initiated a lively discussion within the high-resolution spectroscopic community about the scientific priorities for E-ELT HIRES. In September 2012 a workshop was held in Cambridge to give these discussions a focus point. The workshop was very successful and attracted nearly 100 participants. This White Paper reflects these ongoing discussions, which have also fed into the recent E-ELT instrumentation workshop held in February 2013 in Munich. A major result of these discussions is that the CODEX and SIMPLE concepts are widely considered as rather specialized and that wide instantaneous wavelength coverage and a modular concept would best support a number of truly outstanding science cases, as well as provide the large and vibrant ESO high-resolution community with the high-resolution capabilities they need to be competitive in the E-ELT era.

This White Paper summarizes the science cases and the associated instrument requirements discussed during these meetings, but it also includes the contribution of scientists who could not directly attend.

# 2. Exoplanets

## 2.1. Introduction

The discovery almost 20 years ago of the first giant planet outside of the solar system (Mayor & Queloz 1995) spawned a real revolution in astronomy. The completely unexpected characteristics of this first planet captured the imagination and interest of the scientific community and the general public, arguably like no other result since the discovery of the expansion of the universe. Considered yesterday by most as a wild dream, the study of planets outside the solar system has become reality. Since the first discoveries we came to know more than 800 exoplanets and 2300 additional transiting candidates provided by the Kepler satellite (Batalha et al. 2013). This remarkable outcome is the result of a comprehensive world-wide effort to the goal of advancing our understanding of their formation, evolution, and habitability of planet in the Universe.

It is now clearly demonstrated that planets are common objects orbiting stars in the Universe (Mayor et al. 2011, Howard et al. 2012). Emerging results suggest that a large fraction of solar-type stars host hot Neptune or super-Earth mass planets. These results are supported by the most recent planet formation models (Mordasini et al. 2012). Moreover, the frequency of multiple planetary systems seems to be very large, and most of smaller planets are observed in multi-planetary systems

These unexpected discoveries have sparked a flurry of theoretical activities aimed at understanding the growth of planets, their interactions with the gaseous disc in which they are embedded and with other growing bodies. While the search for solar system analogues is still on going, the first spectra of exoplanets have been taken, signalling the shift from an era of discovery to one of physical and chemical characterization. This will eventually lead to the remote analysis of planet atmospheres with the ultimate goal of life detection.

The bulk of the exoplanets detected by Doppler surveys are gas giants orbiting solar-type stars. In stark contrast to our solar system giant planets, most of them orbit their host stars at distances of less than about 1 AU, and a large fraction have orbital periods less than 10 days - the so-call hot Jupiter or hot Neptune planets. The first measurements of the density of these planets have revealed that some hot Jupiters are strongly inflated by comparison with our own Jupiter. Although many mechanisms may be proposed to explain the inflated radii (Laughlin et al. 2011), none are as of yet satisfying: those models with clear observational consequences like ongoing dissipation of orbital eccentricity have been ruled out for many systems, and no surviving model yet explains the diversity that is observed. In addition while it is accepted that such planets do not form in-situ, their previous and subsequent evolution is still under debate: where did they form, why did they migrate inwards and what halted their orbital evolution?

Different mechanisms have been considered to reconcile the earlier concepts with the new discoveries, in particular the idea that planets can travel or migrate over large distances during their formation epoch. While migration appears to solve some of the issues, others remain puzzling and may hint to more fundamental problems in our understanding. For example, the migration timescale appears to be quite short, so why have not all the planets "fallen" into their star? Why is it that Jupiter appears not to have migrated significantly? Why does our solar system appears so different from most systems known to date? Is this just due to current observational limitations or is it true that our system is more the exception than the rule? Recent models have been developed to try to reconcile migration models and, e.g., the present architecture of the Solar System. Some of these models make predictions on the likelihood of the architecture of extrasolar planets, which can be tested by future observations.

The gravitational tug induced by a planet on a lighter star will be higher, making M dwarfs more amenable to planet detections, in particular of low-mass planets, at the edge of current instrumental precision capabilities. Moreover, planets detected on low mass stars at distance that could allow the existence of liquid water (in the so-called habitable zone, Kastings et al. 1993), like for example GJ581cd, have considerably raised the interest of planet searches on such small stars. The sample of M dwarfs searched for planets is yet small, their lower luminosity and higher magnetic activity makes them more difficult targets than earlier thought. But still, M stars may be for long the only kind of stars where we can realistically hope to detect a signature of life in their atmosphere. Recent Doppler search program as well as micro-lensing surveys results suggest that Earth mass planet at habitable-zone distance orbit may be common for M dwarfs (Bonfils et al. 2013).

In recent years, both space and ground-based observations have established a firm connection between metals in cool white dwarfs and closely orbiting dust disks. A compelling model to account for the circumstellar dust at metal-enriched white dwarfs is the tidal destruction and subsequent accretion of one or more large asteroids or possibly planets (Farihi et al. 2009). The circumstellar material gradually falls onto the star and can be directly observed in the photosphere to obtain its elemental composition in the same way asteroids (via their meteorite fragments that fall to Earth) yield the best available data on the bulk composition of the Earth and terrestrial planets of the Solar System.

Transiting systems are a special and important category of planets, useful to understand the diversity of nature of planetary population in our Galaxy. The special orbital geometry of these systems allows us to measure not only their masses but also their radii, and gather insights on their inner structure and mean bulk composition. This is particularly important for Neptune-mass objects for which a wide range of planetary structures can lead to very different radii. Equally important, we can study the atmospheric properties of transiting planets without needing to resolve spatially the planet from the host star.

By using different techniques and approaches, one can obtain insights on vertical pressure-temperature profile, albedo, heat transport efficiency, circulation pattern, and composition of the upper layers of their atmospheres. By observing an occultation, when a planet passes behind its star, we can measure the *emergent spectra* (or planet's dayside atmosphere) from the planetary atmosphere. Similarly, observation of the planet during the transit at different wavelengths allows us as well to measure its *transmission spectra*.

A rotating star has a blue-shifted hemisphere, and a red-shifted hemisphere. When a planet transits this star, it hides a part of that doppler-shifted velocity. This creates an anomaly in the radial-velocity that we measure from the star. By comparing the time spent over one hemisphere compared to the other, we can obtain a measure of the angle between the planet's orbital spin and the stellar spin projected onto the sky. This technique has been used extensively in the past to study binary systems (Rossiter 1924, MacLaughlin 1924). About 70 Hot Jupiters have had their projected spin-orbit angle measured this way. Almost half of them have been found on orbits that are significantly not coplanar with their star's equatorial plane, including a number on retrograde orbits. The discovery of so many "misaligned" orbiting planets has had profound consequences on our understanding of the formation mechanics. It is has been proposed that those planets undergo dynamical interactions, that leaves them onto highly inclined, highly eccentric orbits. Tides raised during periastron passage eventually circularize the orbit to their current configuration. This scenario, however, would lead to an equal fraction of prograde and retrograde extrasolar planets, a prediction that can be tested observationally.

The study of tidal interactions is interesting in its own right, but tidal interactions also reshape in a non-obvious way the spin-orbit distribution and thus prevent a robust comparison between theoretical models and observations. This is why the emphasis is now placed on smaller mass objects since they raise weaker tides, and longer period objects. Neptune mass planets are particularly interesting since they come both in solitary form like the vast majority of hot Jupiters, but also in multi-planetary systems. One can therefore compare both populations and check whether loneliness is marker of dynamical interactions. Unfortunately, Neptunes also produce much weakened Rossiter-McLaughlin effects because their transits are much shallower. Going into a tidal-weak regime therefore means observing smaller signals. Higher precision Doppler measurement would as well open the door to obtain direct measurement of the differential rotation profile on other stars than the Sun, for which virtually no information currently exist, and whose effects are not included in stellar evolution models.

During a planetary transit, a small fraction of the stellar light is filtered through the atmospheric limb of the planet. The opacity of the upper atmosphere varying with wavelength, the outcome is a wavelength-dependence of the transit depth. Gathering precise transit depths at several wavelengths makes thus possible to measure the transmission spectrum of the planet's atmospheric limb, bringing important constraints on its thermal structure, cloud coverage, and the mixing ratios of the most spectroscopically active among its atomic and molecular components. This is the method that was first used to obtain with HST the first detection of an exoplanet atmosphere (Charbonneau et al. 2002). Since then, the same method has been applied to probe the atmospheric properties of other planets. Notable results were the detection of the exosphere of HD209458b, the detection of a haze of sub-micron particles in the upper atmosphere of HD189733b, and the still questioned detection of water and methane for the same HD189733b. The application of the method has also been so far extended with some successes to two lower-mass short-period planets, namely the Neptune-sized planets GJ436b and GJ1214b that both transit nearby M-dwarfs (Bean et al. 2010).

At high dispersion, molecular bands, either observed in a planet's transmission spectrum or dayside emission spectrum, are resolved in tens or hundreds of individual lines. By observing the combined signal from these lines, these measurements are sensitive to strong Doppler shifts due to the orbital motion of the planet. While the reflex motion of a star around the centre of gravity of a planetary system is typically smaller than a few hundred meter/sec, the planet orbital motion can be up to 150 km/sec. This means that within a few hours of observation, a signal from an exoplanet atmosphere can shift by tens of pixels on the detector, allowing it to be distinguished from both the stationary telluric and the quasi-stationary stellar lines.

In a general sense at the superior conjunction, a transiting planet is occulted by its host star. In the infrared, such occultations make possible the measurement of the light emitted by the planet. The wavelength dependence of the measured fluxes makes possible to reconstruct the Spectral Energy Distribution of the planetary dayside. In addition, occultation timings constrain thoroughly the orbital eccentricity, a key parameter in the accurate determination of stellar and planetary dimensions and in the assessment of a planet's tidal energy budget and tidal history.

The *Spitzer Space Telescope* has produced series of such planetary emission measurements for many hot Jupiters, all at wavelengths longer than 3.5 μm. Recently large-aperture ground-based telescopes, like the VLT, have brought complementary measurements at shorter wavelengths. All together, these measurements have led to detailed constraints on several aspects of hot Jupiter atmospheres, such as evidence for thermal inversions in the dayside atmospheres or detection of non-equilibrium chemistry. The constraints on atmospheric properties of these planets made possible by emission measurements have led to major theoretical advancements in the understanding of irradiated giant planets (Barman et al. 2001). Early theoretical studies suggested two classes of hot Jupiters based on their degree of irradiation (Fortney et al. 2008). The hotter class was predicted to host thermal inversions in their atmospheres due to the possibility of gaseous TiO in their atmospheres, where as the cooler class was predicted to be devoid of thermal inversions. While thermal inversions have been detected in several highly irradiated hot Jupiters, subsequent data have revealed several exceptions to the TiO hypothesis. Consecutive theoretical studies have suggested that non-equilibrium settling may deplete TiO over a wide range in irradiation, and that the inferences of thermal inversions are highly correlated with the chemical compositions.

## 2.2. The characterization of exoplanet atmospheres

Observations of exoplanet atmospheres through transit require overcoming two major obstacles: photon noise limitation and systematics hampering spectral fidelity. From ground the unpredictable variability of Earth's atmosphere, which induces photometric variations much larger than the exoplanet signal and superimposes telluric lines on top of every science observation is an additional challenge to address. At low-resolution lines overlap and confusion between different molecular signatures limits the identification of the planet signal to a few large features. The best way of circumventing these issues is to work at very high spectral resolution to see "through the Earth atmosphere". At high spectral resolution, absorption lines from molecular bands can be detected individually. Planetary, stellar and telluric spectral features can be separated from each other thanks to their differential radial velocities shift. Most importantly, high resolution spectroscopy, by resolving the narrow molecular features and the cores of atomic lines like Na and K can probe a much larger range of altitude in the exoplanet atmosphere. In some case to overcome potential photon noise limitation from working at high resolution, signal averaging technics like cross-correlation may be applied to extract the signal from molecular bands.

As of today, results from high-resolution, ground-based transmission spectroscopy have been obtained both in the visible and the near-IR. In the visible, the most amenable detections are those of alkali metals in the atmospheres of hot Jupiters. In the near-IR, carbon monoxide has been detected in the atmosphere of HD209458b thanks to VLT CRIRES observations (Fig.1, Snellen et al. 2010), while recent works were able to rule out several compositions for the atmosphere of the large super-Earth GJ1214b.

The strong increase in signal-to-noise for observations with HIRES at the E-ELT compared to CRIRES at the VLT will mean that for the brightest systems, not only an ensemble signal from many individual lines can be measured, but that a real high-resolution planet spectrum can be constructed revealing the strengths of individual lines in a molecular band. The lines are produced at a range of different atmospheric altitudes and pressures, meaning that they probe the planet atmospheric temperature-pressure profile in a unique way, with only the molecular abundance as a free parameter. For example, the presence of a thermal inversion layer at a certain altitude range within the planet's photosphere would show up unambiguously as a set of molecular emission lines.

A crucial and fascinating aspect of planetary atmospheric physics is the global circulation or "weather pattern" on a planet, in particular in cases where the dayside hemisphere is strongly irradiated by the star. It governs the level at which stellar energy that is absorbed in the planet dayside hemisphere is transported to the night-side. High-resolution spectroscopy is a unique tool to reveal global wind patterns. CRIRES transmission spectroscopy of the hot Jupiter HD209458b targeting carbon monoxide has shown a signal that appears to be blue-shifted by ~ 2 km/sec, which can be understood as the outcome of high-altitude wind blowing from the hot dayside to the cooler night-side (Fig.1, Snellen et al. 2010). HIRES observations will be able to reveal detailed signatures of planet rotation and circulation, which show up as line broadening, and velocity-shifts during ingress and egress of a transit, when only part of the planet atmosphere is illuminated by the star. This will show, for example, to what extent close-in planets are tidally locked or not, an important aspect to address life issue for smaller planets.

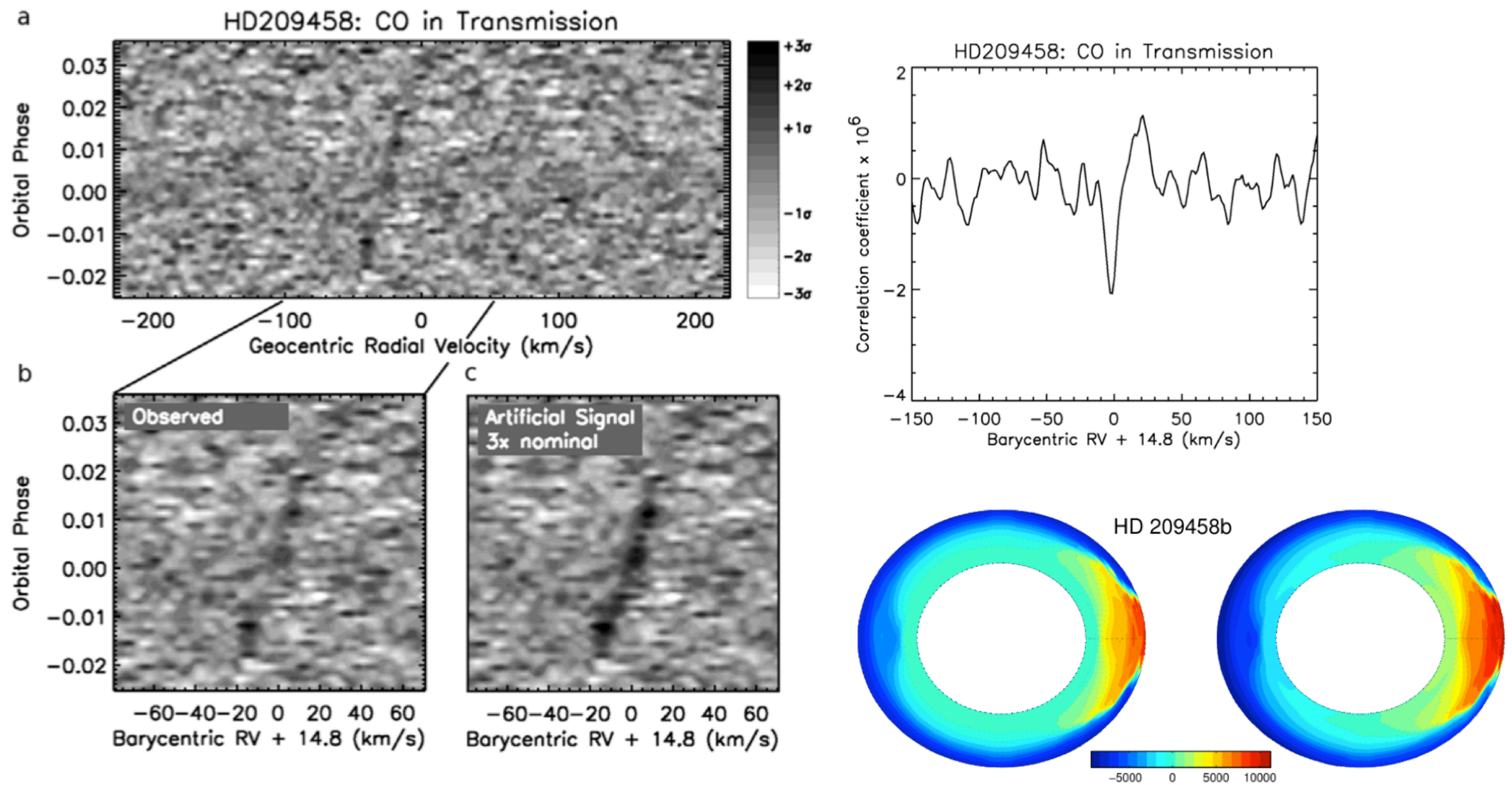

Fig.1. Carbon monoxide signal in the transmission spectrum of exoplanet HD209458b obtained from the cross-correlation of the VLT-CRIRES spectrum with a template spectrum of 56 CO lines. In the left panel the signal is shown as a function of orbital phase (transit occurs in the orbital phase range $-0.02<\theta<0.02$). The top-right panel shows the signal integrated during the transit. The CO-signal is blueshifted by ~2 km/s with respect to the systemic velocity of the host star, which suggests a velocity-flow from the day side to the night side driven by the large incident heat flux on the day side (see Snellen et al. 2010 for details). The bottom-right panel shows a model of the winds radial velocity on the same planet, from Showman et al. (2012).

Another important aspect of high-resolution dayside spectroscopy is that a planet signal can be retrieved along a large part of its orbit, not necessarily directly around secondary eclipse. This means that molecular signals can be obtained as function of planet longitude, c.f. from the morning-side to the evening-side of the planet, revealing possible photochemical processes or variations in the atmospheric temperature structure. In addition, HIRES observations of the brightest systems may reveal other molecular isotopologues, providing insights in the atmospheric evolution of exoplanets, e.g. through evaporation.

The strength of the transmission signal of an exoplanet atmosphere depends on the star and planet sizes, as well as the scale height of the atmosphere. The scale height itself depends on the mean molecular mass, which varies by one order of magnitude between a hydrogen-dominated and hydrogen-depleted atmosphere. Therefore, there is a large variety of possible signal amplitudes depending on the exoplanet properties. Quantitatively, these range from a few thousands of ppm for gaseous giant planets to only a few ppm for an Earth-like planet. These numbers show how challenging these observations are, both in terms of SNR requirements and control of systematic effects.

A signal-to-noise ratio per resolving element above thousand is a minimum to detect a transmission signature of a planet. Likewise signal-to-noise ratio from 10,000 to 1,000,000 should be the goal to study Neptuns and study more features on giant planets. Practically, it means that up to $10^{12}$ photons must be collected during primary transit, i.e. within a few hours of observations. The ability to achieve this in a single transit is an important aspect to minimize systematic effects from the instrument and from stellar activity, as well as an optimal way to catch time-critical transit events. This unprecedented number of photon that needs to be collected, in apparent contrast with the observation of the most distant objects usually done on big telescopes, illustrates an important twist in ground based astronomy. *In a apparent paradox we are in absolute need of very large telescope apertures to study the most interesting exoplanets, with a compact atmosphere on nearby bright stars!* This reality brings upfront the competitive asset of the ELT against space experiment, where mirror size is limited. With its 39-m diameter, the E-ELT has a unique role to play in this field.

Simulations, extrapolating from high resolution spectra observed with HARPS, show that molecular band can be detected in bright stars with known exoplanets in transit, including Neptuns and "super-Earths" such as GJ3470 b, GJ1214 b and 55 Cnc e. Fig.2 shows that, through the cross-correlation technique, HIRES will even be able to detect the signature of $O_2$ in Earth-twins transiting M5 dwarfs. As more exoplanets orbiting bright stars are discovered, the diversity of exoplanets within the reach of E-ELT HIRES will increase, eventually yielding a comprehensive view of atmospheric properties of exoplanets, from hot Jupiters to temperate super-Earth exoplanets.

Finally, very interesting information can be gathered about the exoplanet atmospheres if we are able to detect the optical reflected light spectrum of the star on the stellar disk. Such detections, while challenging with present day instrumentation, will be possible for a large number of known planets with HIRES at the ELT. Measuring the reflected light spectrum will be possible even in cases where planets do not transit, and will allow to determine the orbital velocity of the planet and thus its mass, like in the case of stellar double-line eclipsing. Furthermore, it will give critical information about the atmosphere physics of the planet (e.g. geometric albedo), or even about its rotation rate.

The amount of reflected light that we can expect to detect depends on different physical quantities: it is a function of the planet radius, its distance from the star, the planetary albedo, and the percentage of the planet disk that is illuminated (i.e. its phase). The first two points imply that short period giant planets (hot Jupiters) are the best candidates for such a detection. Exoplanetary albedos are more difficult to estimate, but values may reach (or even exceed) 0.3. In any case, the expected signal is usually very small (Flux_planet/Flux_star) around $10^{-4}$ to $10^{-5}$ for a hot Jupiter, and one order of magnitude smaller for Neptunes and Super-Earths. The large aperture of the E-ELT, as well as a high fidelity high resolution spectrograph, will thus be of prime importance for this case. More specifically, very recent simulations show that such reflected light studies are well within reach of HIRES at the E-ELT (Martins et al. 2013).

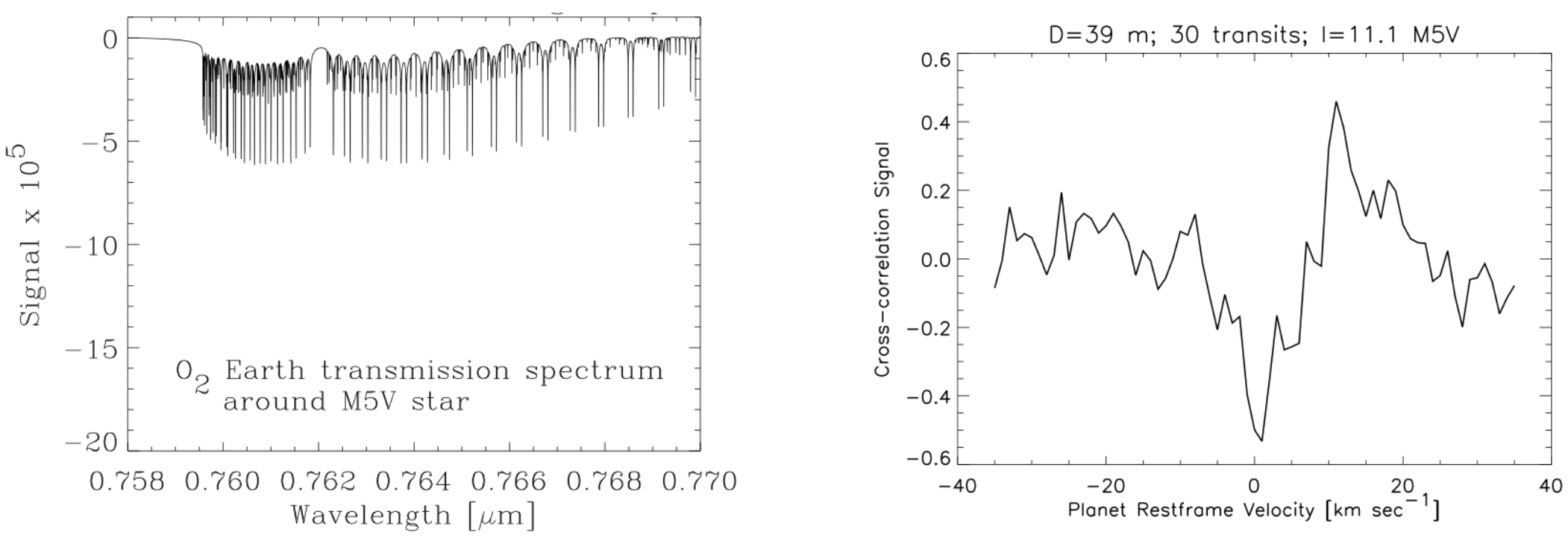

Fig.2. *Left:* Simulated $O_2$ transmission signal for an Earth-twin transiting an M5 dwarf star. *Right:* Simulated $O_2$ cross-correlation signal from a hypothetical Earth-twin in the GJ1214 system, showing a 3.8 σ detection after collecting data for 30 transits with the 39m E- ELT, with HIRES at R=100,000 and assuming a total throughput of 20%. From Snellen et al. (2013).

Summarising, the HIRES instrumental requirements related to characterization of exoplanet atmospheres are:

- Spectral resolution 100,000 or higher. This is absolutely needed to separate telluric and exoplanet line systems, to maximize the contrast of narrow molecular lines (hence probing different atmospheric layers), as well as to be sensitive to planet rotation and atmospheric circulation.
- Wavelength coverage: 0.38-2.40 μm. There are potentially interesting spectral features almost everywhere in the visible/near-IR range. The blue limit is set by the reflected light case, since the geometric albedo is expected to increase towards the blue (Evans et al. 2013). The inclusion of the (blue) Ca II H&K lines for monitoring of stellar activity is also relevant, while the inclusion of the K-band is motivated by the presence of a strong CO band at 2.3 μm.

- High stability of the PSF and spectrum position on the detector, of the order of 1/100$^{th}$ of a pixel. This also implies a stability of the wavelength calibration of 10 m s$^{-1}$ during the transit. This points towards a stabilized fiber-fed instrument in a stable, gravity-invariant environment, with a fixed spectral format and no moving parts inside the spectrograph (like HARPS and ESPRESSO).
- Good detector stability and high signal level capability: high-precision flat-fielding (better than 0.1%), negligible gain/sensitivity variations within a night. Precise calibration of detector non-linearity.
- Adaptive optics is not required. This is a "point-source" science case with no need of spatial resolution.
- Multiplexing is only useful if differential spectrophotometry is possible. Probably not at R=100,000 (variable slit losses), but maybe at lower resolution. Still, a simultaneous telluric spectrum obtained using a nearby hot star would be very useful if practical.

## 2.3. Radial Velocity Follow-up of terrestrial transiting planets

Both *Kepler* and radial velocity (RV) surveys have shown that small-size ($r \leq 5\ R_\oplus$) and low-mass ($m \leq 30\ M_\oplus$) planets are numerous, but the determination of their true density is far to be trivial. The relative lack of such low-mass planets in the mass-radius diagram is due to the current detection limits. Furthermore the uncertainties on their mass and radius unfortunately do not allow us to properly constrain their internal structure nor composition. This situation mainly comes from the fact that high-precision photometric surveys like *CoRoT* and *Kepler* observe quite faint stars ($m_V$=12-16), which are not appropriate for high-precision radial velocity follow-up. The next generation transit surveys (NGTS, TESS, PLATO) will focus towards brighter stars ($m_V$=10-13). One may estimate that several hundreds of confirmed low-mass exoplanets will be known in the next decade. However, the characterization of their mass and radius with an accuracy better then 10% is crucial to study their internal structure and composition and to distinguish iron-rich, silicate, icy, and mini-Neptune like exoplanets. The mass characterization with 10% accuracy of terrestrial planets ($m \leq 10\ M_\oplus$) requires RV precision measurements better than 30 cm s$^{-1}$. This level of wavelength calibration accuracy and the sensitivity of HIRES at the ELT will allow us to measure the mass of validated transiting terrestrial planets discovered from the next generation transit surveys ($m_V$=10-13) with uncertainty of less than 10%.

The current push towards temperate planets (orbiting on longer period), and small planets calls for the capacity to detect weak, and rare and time limited signal like the Rossiter-McLaughlin effect. The observation of the shadow effect of a small planet by Doppler techniques requires the capability to obtain precise measurements with high cadence. The Rossiter-McLaughlin effect from a transiting Earth-analogue would make a 16 cm s$^{-1}$ radial-velocity "anomaly", slightly larger than the gravity pull effect on the star 9 cm s$^{-1}$. But this signal would last for 10 to 12 hours, making the confirmation of Earth-like transiting planets matter of a couple of nights instead of several years of observing efforts. If it is possible to obtain the Doppler shadow signature of an Earth like planet, it will also be possible to detect similar sized objects in orbit around long period transiting gas giants. The large moons in the solar system are closely aligned with the equatorial plane of their planet, even in the odd case of Uranus, hinting that such campaigns might indeed deliver interesting results.

## 2.4. Chemical Characterization of Planetary Debris

Insights on composition of rocky planetesimals and possibly planets themselves can be gained by spectroscopic observations of metal-enriched white dwarfs. Critical elements list for a detailed, chemical analysis of the planetary debris on these objects includes:

- the major rock-forming elements (O, Mg, Si, Fe);
- the volatiles which are depleted in rocky material that underwent thermal processing (C, N, Na)
- the elements indicative of differentiation (Cr, Fe, Ni)
- refractory lithophiles (Al, Ca, Ti)
- tracers of water (H, O).

Observations of lines related to these elements requires high resolution and high signal-to-noise ratio. The current state of the art facilities on 8m class telescopes are limited a to dozen of bright metal-polluted white dwarfs, which have already delivered extremely exciting results, such as the identification of planetary material with composition similar to the Earth-Moon system (Fig.3). With *GAIA* and massively multiplexed ground-based spectroscopic with 8-m class telescope, several fainter white dwarfs (up to 19$^{th}$ magnitude), of which a few hundreds "polluted" by planetary debris will be discovered. They will be ideal targets for further characterization with the E-ELT. Therefore, an efficient, high resolution blue-sensitive spectrograph will be a key instrument to carry out a detailed statistical analysis of the bulk chemistry of building blocks of planetary systems.

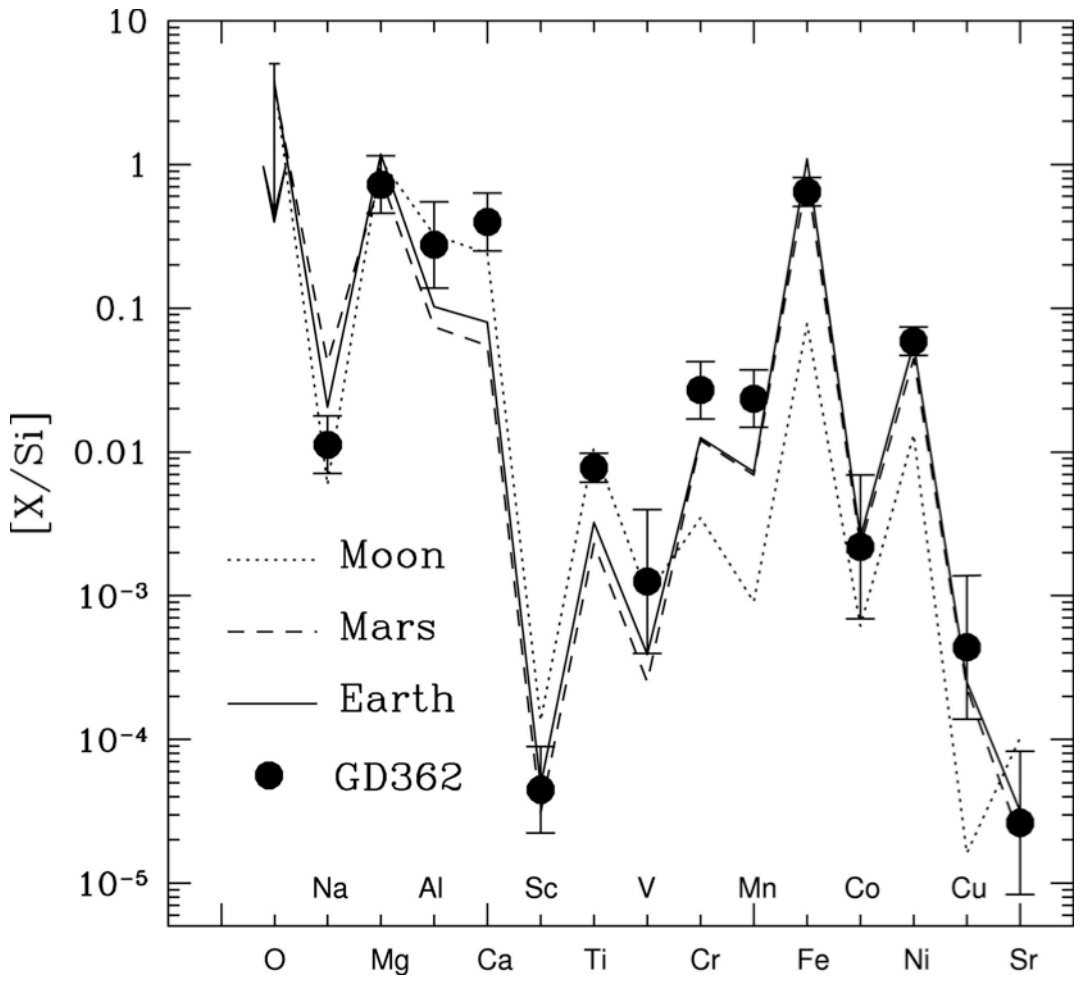

Fig.3. A spectacular example of the measured composition of the infalling planetary material on the white dwarf GD362, thanks to high resolution spectroscopy. The destroyed parent body has a basic composition similar to the Earth-Moon system (Zuckerman et al. 2007).

## 2.5. Potential of the Spectro-polarimetry

The combination of a dual-beam polarimeter with a high resolution spectrograph not only allows to observe fundamental effects like the Zeeman and Hanle effects, but also permits a refinement of the spatial resolution within the plane of the sky and a radial dimension to be added where zones with different physical conditions. A transiting planet breaks any spherical symmetry over the projected stellar disk and thus results in a non-vanishing linear polarization. In the case of an unresolved star-planet system, the degree of polarization being essentially given by the ratio of reflected to total (i.e., polarized + unpolarized) light is expected to be small. However, when we resolve the solar disk, a polarization variation from center to limb occurs due to electron (Thompson) scattering in hot stars and Rayleigh scattering by atomic and molecular hydrogen and atomic helium in cool stars. This may be detectable on other stars. Resonance scattering excited by the stellar radiation is the source for spectral polarization features in the light of molecular bands of a planet's atmosphere. Both effects together, center-to-limb variations from the stellar disk and enhancements of molecular line polarization, require high precision linear polarization measurements at high spectral and spatial resolution. Circular polarization has been detected in the spectrum of the Earthshine just recently with the VLT and the FORS2 spectro-polarimeter. It is interpreted as a signature of biomatter (chlorophyll in particular) on the Earth's surface.

Complex organic life on Earth may not have developed if the magnetic field of the Sun and Earth did not protect the Earth from cosmic rays. Recent studies have found enhancements in magnetic-activity indicators in planet-hosting stars that vary with the orbital period of the planet, suggesting that some extra-solar planets may indeed possess magnetic fields. But these perturbations may also be of a different nature and ultimately require a direct detection. High-resolution spectro-polarimetry with the E-ELT may open the path to such a first detection and thereby contribute to the search for extraterrestrial life.

**Table.1.** Summary of science requirements for the main **exoplanet** science cases (**E**=essential; **D**=desirable)

| Science case | | Spectral resolution (λ/Δλ) | Wavel. range (μm) | Wavel. accuracy (cm $s^{-1}$) | Stability (cm $s^{-1}$) | Multi-plex | Backgr. subtr. | AO / IFU | Polarim. |
|---|---|---|---|---|---|---|---|---|---|
| Exoplanet atmospheres | **E** | 100,000 | 0.38-2.4 | 1000 (relative) | 10 $night^{-1}$ PSF+detector <0.1% | none | not crit. | no | no |
| | **D** | 150,000 | 0.38-2.4 | 100 (relative) | 5 $night^{-1}$ PSF+detector <0.01% | 2[a] | not crit. | no | yes |
| Planetary debris on White Dwarfs | **E** | 100,000 | 0.37-0.5 | not critical | not critical | none | not crit. | no | no |
| | **D** | 100,000 | 0.33-0.5 | not critical | not critical | none | not crit. | no | no |
| Orbital elements | **E** | 100,000 | 0.4-0.9 | 10 | 10 | none | not crit. | no | no |
| | **D** | 150,000 | 0.4-0.9 | 10 | 10 | none | not. crit. | no | no |

Notes: [a] For telluric calibration.

# 3. Stellar structure and evolution: inside-out

## 3.1. Introduction

In the ALMA and JWST era, an E-ELT class telescope with a high resolution spectrograph, covering the full optical and infrared (IR) spectral range between 0.37 and 2.5 μm, represents a major opportunity for new, fundamental knowledge in stellar astronomy and astrophysics.

Stars are far from being understood. In particular, dynamical phenomena and phases of stellar evolution are not well explored. Although new theoretical simulations of stellar convection, differential rotation, stellar activity, mass loss, and interior mixing, as well as star formation and the interaction of stars with protoplanetary disks and planets, are being developed, these efforts are to a considerable degree depending on simplifications and unknown initial conditions, and on free parameters. This primitive state of our knowledge leads to severe limitations of using stars as probes in the empirical study of the evolution of planetary systems and galaxies, and well in cosmology. The only way to remedy this situation is to guide theoretical developments using better observations. With more sensitive instrumentation, it will be possible to dramatically improve our understanding by testing theories using stars of lower mass, with different chemical composition and age, than those available for study with present telescopes. This will increase the value of the use of stars in various applications in almost all fields of astronomy. In many cases, this increase may become of fundamental significance.

When compared to the current generation of ground-based 8-10m class telescopes, the considerably larger aperture of an E-ELT leads to a substantial gain in sensitivity. With an E-ELT and an efficient high resolution spectrometer it will be possible to reach stars of $19^{th}$ magnitude over the full optical and near IR range, at high spectral resolution (R>100,000) and high S/N (>100), i.e. to obtain spectra of this quality for solar-type and cooler dwarf stars out to distances of several kpc, thus sampling most of the Galactic disk and bulge, or sub-giants and red giants in the outer Galactic Halo and in the neighbouring dwarf galaxies.

Fig.4 summarizes some of the science cases enabled by the E-ELT, with respect to the state of the art with 8-10m telescopes, discussed in this section (but also cases that will be presented in the coming sections), by also showing some HIRES simulated spectra, compared with results obtained with the instrumentation available so far.

This is a major improvement to our current knowledge based on studies in the solar neighbourhood since much more distant regions can be reached without being dependent on spectra of super-giants and bright giants, that are notoriously difficult to analyse due to the low gravity and complex dynamics in the stellar atmospheres. It will be possible to increase the accuracy in spectroscopic chemical analysis by at least one order of magnitude, thus enabling a detailed determination of the "chemical-element profile" of stars rather than a rough chemical tagging. Currently, such accuracy is only available for a few dozens of stars.

The value of such new/extended knowledge has a considerable potential to provide truly revolutionary results within several disciplines of astrophysics, such as star and planet formation, stellar physics, star and galaxy evolution, nucleosynthesis, and chemical evolution.

The new knowledge will mostly stem from three main actions, namely (1) enabling new stellar physics observations, (2) probing stellar populations in new environments, (3) finding new and important fine structure patterns among stellar populations and (4) unveiling the unexplored.

However, in order to reach these goal, it is fundamentally important that the spectra are not only nominally of high resolution and have high S/N (e.g. Dravins 2008), but they must also cover the widest possible spectral range from the optical to the near IR and be very well calibrated (in terms of wavelength, flux, flat-fields, etc.).

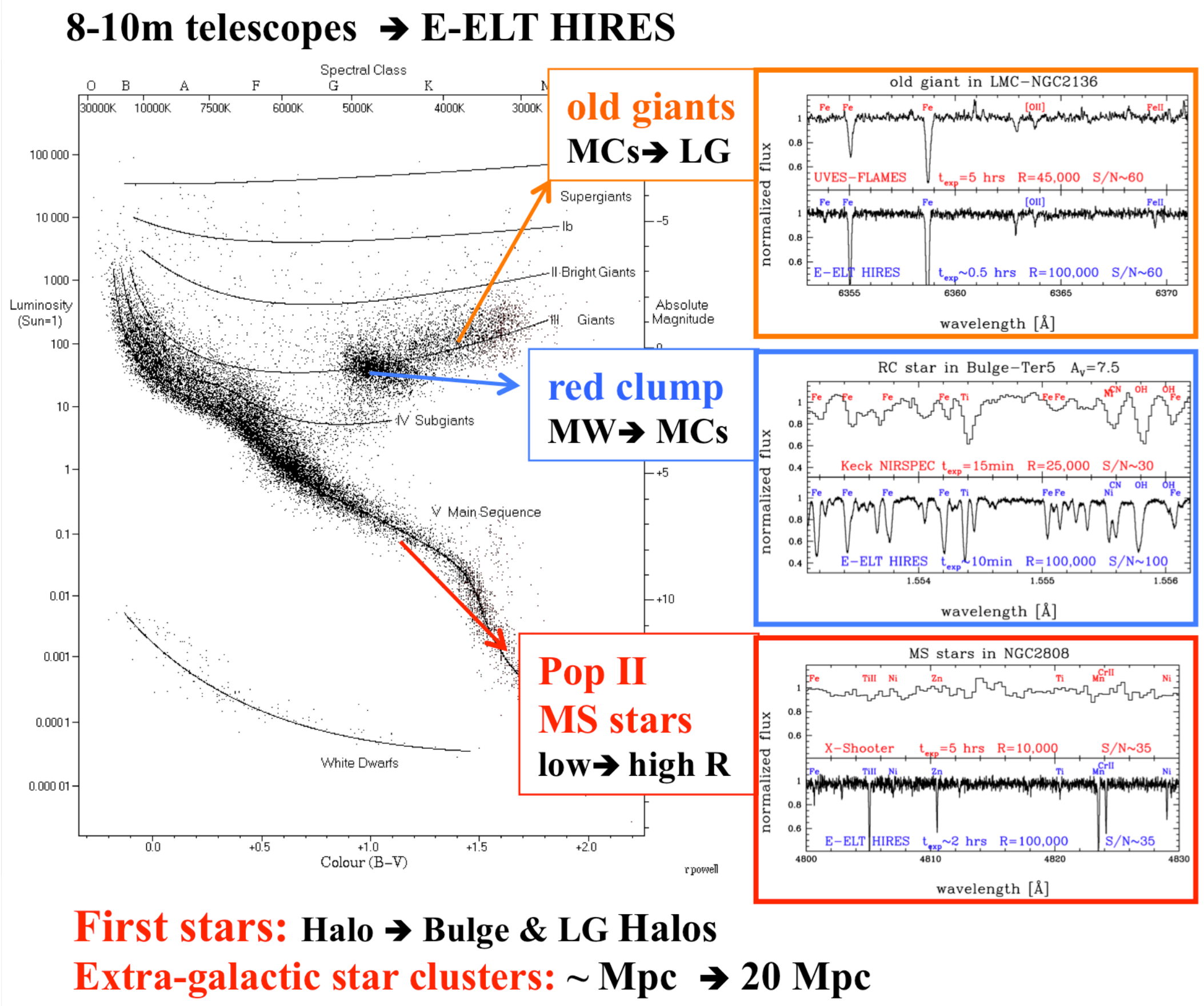

Fig.4. Diagram sketching some of the discovery space of HIRES in the field of stellar astrophysics with respect to instruments at 8-10m telescopes. HIRES will enable to expand the investigation of old giants from the Magellanic Clouds to galaxies in the Local Group and will enable to extend the investigation of red clump stars from the Milky Way to the Magellanic Clouds. HIRES will allow us to investigate population II Main Sequence stars at high resolution and will allow us to study the first, very metal poor stars in the bulge and in the halo of Local Group galaxies. Finally HIRES will enable the investigation of extragalatic star clusters out to a distance of 20 Mpc, i.e. including galaxies in the Virgo cluster.

### 3.2. Stellar atmospheres

The quantitative study of stellar atmospheres is of primary importance to understand the 3D structure of stars, the production of thermonuclear energy and the nucleosynthesis of metals, with major impact in Galactic, stellar, and exo-planet research. Stellar effective temperature, surface gravity, radius, and detailed chemical composition must be quantified with great accuracy.

For many stars, stellar surface gravities and temperatures are normally derived from excitation and ionization equilibrium of weak spectral lines of different species. Valuable information can be added by the wings of strong lines (e.g. Balmer lines) and, in many cases, this information proves essential (e.g. Nieva & Przybilla, 2012; Fuhrmann et al., 1993, 1997). The precise recovery of line profiles requires high S/N at high spectral resolution, as offered by an ELT aperture. This is also essential for the analysis of convective motions from weak photospheric lines. Absorption lines formed in convective atmospheres become asymmetric and usually blueshifted in wavelength due to the asymmetric contributions of light from hot, bright, and rising (thus locally blueshifted) granules and from cooler, darker, and sinking (thus locally redshifted) inter-granular lanes. The 3D-structure of convective motions changes with depth, and it is a property that today is very difficult to measure, but possible with a suitable E-ELT HIRES facility. A measurement of line profiles with an accuracy in the rest intensity better than 0.5% is needed and relative flux levels in the wings of strong lines must also be recovered to this accuracy (e.g., Ammler-von Eiff & Santos 2008), while the cores of important strong lines (Hα, Hβ, MgIb) should be located close to the center of an Echelle order.

The surface chemical abundances of stars reflect the initial composition of the gas from which they formed but also possible modifications occurring during their evolution, as a consequence of different processes such as atomic diffusion, dredge-up and other mixing mechanisms. First, accurate studies of abundances along the Main Sequence, Sub-Giant and Giant Branches of globular clusters by Korn et al. (2007), Lind et al. (2009) and Nordlander et al. (2012) have suggested the prensence of atomic diffusion, leading to individual depletion signatures of metals like Mg, Ca, Ti and Fe of the order of 0.05 to 0.2 dex for stars at the turn-off point, as compared with the giants.

Recent results suggest that similar effects are also active for more metal-rich stars. These effects contain important information on the evolution of the stellar surface composition, which can be used for dating of stars, as well as for studies of their interior dynamics. They must be considered when the chemical evolution of galaxies is explored in detail. Systematic studies of diffusion in globular cluster stars need, however, an ELT.

Studies of the CNO processed material onto the surfaces of red-giant stars (diminished C, increased N, decreased $^{12}C/^{13}C$ ratio) during the first ascent along the giant branch have highlighted the existence of the dredge up phenomenon , but still the theory lacks a full agreement with the observations and extra mixing processes have been proposed (see Angelou et al. 2011 for a summary), including thermohaline mixing (see, e.g. Charbonnel & Zahn, 2007). In order to verify this, or to analyse other possible mechanisms, it is mandatory to get more accurate C, N and O abundances (better than the 0.1 dex achievable today) from CN, CO, and $C_2$, and it is also indispensable to determine isotopic ratios (e.g. $^{13}C$, $^{17}O$ and $^{18}O$ isotopes from $^{13}C^{16}O$, $^{12}C^{17}O$ and $^{12}C^{18}O$). A study of the phenomenon along the entire extension of the main-sequence, sub-giant and giant branches of globular clusters of different metallicities is critical to this purpose and it requires an E-ELT aperture to measure these stars, in particular in Magellanic Cloud clusters. Future combinations of these data with stellar seismology data will reveal the connection between dredge up and differential rotation.

Accurate determinations of stellar parameters will also require the use of 3D hydrodynamic model atmospheres (e.g. Magic et al. 2013) for interpreting the observations. However, except for the Sun, it is difficult to verify or falsify such complex models. One class of predictions involves the changing shapes and wavelength shifts of spectral lines across stellar disks, from disk centers toward their limbs. Although this does not require the fine structure of stellar surfaces to be resolved, some spatial resolution is of very great value. One possible observing strategy is using a transiting exo-planet to successively shadow different parts of the stellar disk. Alternatively, operating near the diffraction limit in the H band, the E-ELT will provide an angular resolution of about 10 mas, which will allow us to measure the center-to-limb variation of the spectrum with a few resolution elements in nearby stars. When such measurements are made for different classes of spectral lines (i.e. with different strengths, excitation and ionization potentials, wavelength, etc.), quite detailed tests of 3D models will become feasible.

*Solar twins.*

First studies of solar-twins and solar-like stars (e.g. Meléndez et al. 2009) seem to indicate that the Sun departs from most twins in the ratio of abundances of volatile elements (e.g. C, N, O, S, Zn) relative to refractories (e.g. Fe, Ni, Al, Sc, Zr) by typically 0.08 dex. This might suggest that the solar proto-planetary nebula was once cleansed from refractories by planetary formation or other processes. Later, the dust-cleansed nebula was accreted by the proto-sun and it may have led to the observed ratio if the convection zone was then not very deep, perhaps as a result of episodic accretion (e.g. Baraffe, Chabrier & Gallardo 2009). This high ratio has been found also in a few other stars (Önehag et al. 2011, Meléndez et al. 2012). If these results are confirmed, a far-reaching implication would be that the surface abundances of solar-type stars may not be totally representative of the composition of the gas/dust cloud from which the star was once formed. Moreover, the effect may contain information on the accretion history of the star. Another possibility is that the solar composition is characteristic of stars formed in a stellar-rich environment. The phenomenon illustrates the unexpected discoveries that tend to follow from raising accuracy in astrophysical measurements significantly. However, observations of larger samples of twins beyond the Solar neighbourhood are mandatory to obtain information on the possible correlation between this solar signature and the existence of terrestrial planets, and also because this issue is still far from being settled, as other literature results do not seem to confirm this situation (e.g. Gonzalez Hernandez et al. 2013, Shuler et al. 2011).

*M-type and ultra-cool dwarfs.*

M dwarfs are the most abundant stars in our Galaxy (>72%) and their precise modeling is very important for 1) understanding fundamental physical problems (magnetism, convection etc.); 2) determining precise (M–L) relations at different metallicities for a correct derivation of their contribution to total mass of the Galaxy and related implications for cosmology; 3) understanding their fundamental properties.

Despite their importance, M dwarfs are red and faint, with optical spectra overcrowded with spectral lines. The derivation of precise parameters for M-dwarfs has thus remained one of the most challenging aspects of stellar spectroscopy. This problem is becoming more and more relevant and important to solve, as (for example) many planet search programs are now concentrating their searches on M-dwarfs: these are the most favorable targets for the detection of an Earth like planet in the Habitable Zone. However, the derivation of precise parameters for exoplanets orbiting these cool stars, including their masses and radii, is critically dependent on the knowledge of the stellar parameters. As such, several teams are now trying to address this problem, and conclude that the derivation of precise parameters for M-dwarfs will be possible by NIR high-resolution spectroscopy, provided that a sufficiently high S/N can be obtained. Indeed, NIR spectra are much less affected by the severe line blending seen in optical spectra of these cool objects (e.g. Önehag et al. 2012).

In this context, HIRES at the E-ELT will offer a unique opportunity to address the determination of precise parameters for thousands of (faint) M-dwarfs, with or without known planetary companions. Furthermore, it will allow the measurement of precise chemical abundances for many different elements (including alpha and iron peak) in M-dwarfs, a step never explored in detail. This study will also open a new window to other fields of research, such as the chemical evolution of the Galaxy.

E-ELT HIRES will be also crucial to study the asteroseismology of M dwarfs. Indeed, the Fourier spectrum of pulsation modes produced by stellar oscillations represents a critical observable to determine the star radius, mass, age and to probe its interiors. By observing classical pulsators (rapidly oscillating delta Scuti stars) and solar-like oscillations in M dwarfs, characterized by the smallest amplitude and fastest pulsations, one can also establish the border line between partially- and fully-convective structures.

Finally, E-ELT HIRES will be the first instrument able to perform quantitative spectroscopy in young (star forming regions, clusters) and old (field) brown dwarfs. This includes element abundances and atmospheric structure analysis as in stars, as well as dust surface distribution and magnetic fields. Understanding the physics of brown dwarf structure and atmospheres is urgently required to put stellar and planetary evolution within a self-consistent picture and to understand the role of magnetism in the formation and evolution of stars and planets.

*AGB stars.*

The AGB stars are major contributors to nucleosynthesis in the galaxies, providing carbon, and affecting oxygen, including their isotopes, and many s-elements. Due to the complexity of the physics involved – with mixing in the interior during and after the thermal pulses, the transport to the surface in these highly dynamical atmospheres, and the final mass loss – the true yields of elements from these stars, as a function of stellar mass, initial chemical composition and initial angular momentum, are quite uncertain although essential progress has been made in recent years (see e,g. Cristallo et al. 2009). The uncertainties still lead to considerable ambiguity in contemporary models of the chemical evolution of galaxies. The situation requires very accurate abundances as functions of the fundamental stellar parameters and in different galaxy environments. For this purpose, AGB stars as well as stars strongly affected by mass transfer from such stars (primarily Ba II stars) in the Local Group of galaxies should be inter-compared in detail, which is not possible today.

All the science cases in Sect. 3.2 require: **1**) high resolution (R>100,000) and very high S/N (>100) spectra for accurate line profile analysis and chemical abundance determination (significantly better than 0.1 dex), **2**) full optical/near IR spectral coverage for complete line diagnostics and physical/chemical information.

Some spectro-polarimetric capabilities to probe magnetic fields in a number of stellar conditions, evolutionary stages and environments are also highly desirable.

An E-ELT aperture is needed to get enough sensitivity and accuracy for quantitative spectroscopy of stellar atmospheres in (**i**) dwarf stars well beyond the Solar neighborhoods and/or cooler than the Sun, as well as in (**ii**) giant stars beyond the Galaxy, thus probing stellar physics in new conditions and exploiting high-precision stellar spectroscopy for studies of different environments.

## 3.3. Early and late stages of stellar evolution

During their lifetime stars change their structure, hence it is of primary importance to understand how stars form and evolve to their final fate. E-ELT HIRES will be a unique facility to perform quantitative spectroscopy of low mass stars in their early (pre-Main Sequence, PMS) and late (white dwarf, WD, cooling sequence) stages, characterized by too low luminosity for observations at high resolution with 8-10m class telescopes.

*The true infancy of proto-stars.*

Our knowledge of the formation history of solar mass stars is mainly based on systematic observations of large populations of PMS stars (the Classical TTauri stars), which are young ($\sim 10^6$ yr) stars still retaining a moderate accretion and jet activity. However, in order to track back the evolution of solar type stars, it is of primary importance to understand how and if models constructed for TTauri work for less evolved systems, still embedded in their original infalling envelope and likely deriving most of their luminosity from accretion through a massive disk (the so-called class I sources, age of $\sim 10^5$ yr). For such sources, the large extinction along with the strong IR excesses due to the circumstellar envelope makes the investigation of the stellar and disk properties extremely challenging. Currently, weak photospheric features and emission lines from the inner disk have been detected in a few class I sources with low/moderate IR excess, only (e.g. Nisini et al. 2005, Antoniucci et al. 2008).

Such observations suggest that not all the embedded Class I proto-stars are indeed accretion dominated, even if their mass accretion rates are on average higher than for more evolved stars of the same mass. Observations in the near IR at R~100,000 are required to detect the weak and narrow absorption lines against the strong continuum in samples of class I sources with different masses and IR excesses, giving unique constraints on the mass accretion evolution during the first stage of star formation.

*Star/disk/planet formation.*

Recent surveys in the mid-IR show that the paradigm of star formation via core-collapse and the formation of an accretion disk, that survives for large part of PMS phase very likely applies to all low-mass objects, down to planetary masses (Natta et al. 2006, Alcala' et al. 2008). However, disk properties (geometry, gas chemistry, dust composition and grain size, accretion and mass loss rates, etc.) show huge and unexplained variations both for stars of similar mass and over the mass spectrum. We are still missing crucial ingredients that control the disk physics; this is particularly important when addressing the issue of planet formation, which depends on disk properties, such as the mass accretion rate, the ejection of jets and winds, and the growth and settling of dust, as well as metallicity. Studies in the Magellanic Clouds (De Marchi et al. 2011, Spezzi et al. 2012) suggest that metal-poor stars accrete at higher rates compared to solar-metallicity stars in Galactic star forming regions. Also, it has been argued on theoretical grounds that the efficiency of dispersal of circumstellar disks depends on stellar metallicity, i.e. the formation of planetesimals around stars may be faster in a high metallicity environment (Ercolano & Clarke 2010). Therefore, simultaneous measurements of accretion rates and metallicity in young stellar objects (YSOs) in different environments are crucial.

Circumstellar disks first mediate accretion onto the central star, triggering energetic collimated jets, and later on, they act as the reservoir of material out of which planetary systems form. However, the interaction between the star-disk system, the jets and the forming planetary systems are very poorly understood and they require broad-band and high resolution observations of the inner disks, where the most active regions are located. Several proto-planetary disks show an inner hole (a fraction to tens of AU) in the dust distribution much larger than its sublimation radius (e.g. Najita et al. 2007; Isella et al. 2006). Possible explanations for these inner holes include dynamical clearing by large bodies orbiting the star (planets or companions), photo-evaporation, grain growth and viscous evolution of the disk. Recent ALMA results revealed that gas flow from the outer disk through the planets and into the central stars are also possible at this stage (e.g. Casassus et al. 2013). This planet-mediated accretion may have a critical impact on the evolution of the inner planetary system and the removal of the outer disk. As the (originally molecular) gas crosses the low optical depth path, it becomes much more exposed to the direct stellar radiation, enabling photodissociation, excitation and evaporation, and optical/IR observations of the warm/hot molecular (e.g. $H_2$, $H_2O$, OH, CO lines in the K band ) and atomic (e.g. the [OI] 630 nm line) gas are thus crucial to probe the physics and kinematics of the inner disk. R>50,000 is needed to resolve the line profiles (<15 km/s), to infer size, geometry and dynamics of the emitting region and to detect weak and narrow $H_2O$ lines against the telluric contribution.

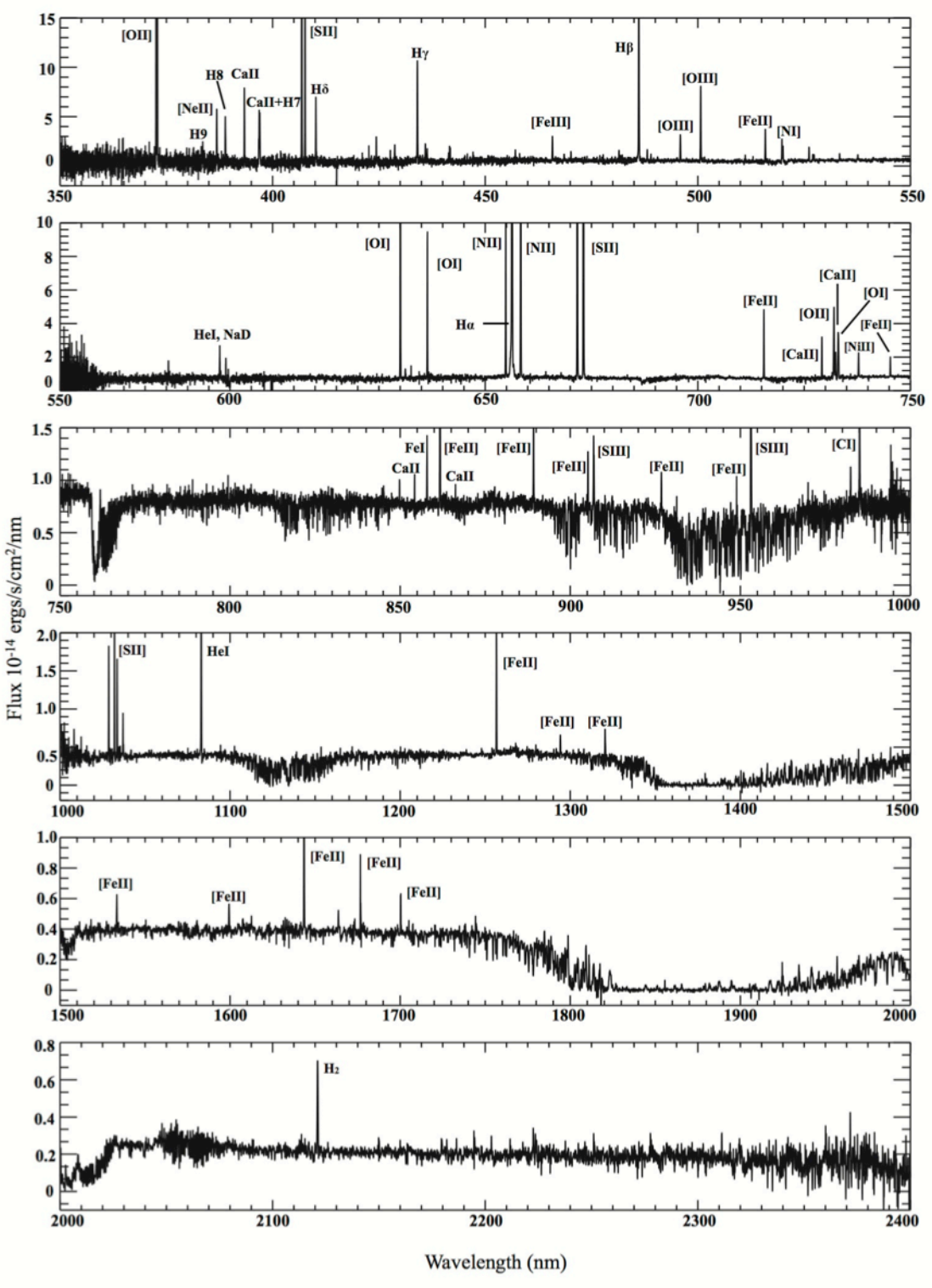

Fig.5. VLT X-shooter spectrum of the jet from a PMS star (ESO-HA 574), showing the benefit of having simultaneous 370-2400 nm spectral coverage at high resolution to measure multiple diagnostic features. From Bacciotti et al. (2011).

It is also very important to distinguish between the various jet launching scenarios (e.g. from the stellar surface, the magnetosphere/disk interface or from the so-called "disk winds", Ferreira, Dougados, & Cabrit 2006), and to constrain the rotation of the star-disk system and the consequent removal of angular momentum by the jet (e.g. Bacciotti et al. 2002). Velocity shifts related to rotational signatures are of the order of few km/s, thus high spectral resolution is needed to detect this effect. As shown in Fig.5, the study of jets will great benefit by the simultaneous access of a large number of forbidden lines and H2 lines, from UV to NIR, able to probe the different excitation layers in the jet beam.

All these regions are mostly spectrally and spatially unresolved with the current or near-future space facilities and AO systems working on 8-10m telescopes (see Carmona 2010 for a review). Indeed, one needs to probe (**i**) spatial scales of ~10 mas, i.e. down to Earth-orbit scales in the nearby star forming regions, and (**ii**) velocity fields in the ~km/s range (typical values of the Keplerian velocities in these regions).

High sensitivity E-ELT HIRES spectroscopy, simultaneously covering (to avoid problems related to time variability) the full 0.38-2.5 μm spectral range will make it possible to: **a**) characterize Galactic YSOs and their disk properties for masses far below the hydrogen burning limit, down to the planetary mass regime; **b**) study mass accretion and ejection activities as a function of mass, age, metallicity and environment by investigating YSOs both in the Galaxy and in the MCs; **c**) probe disk evolution as a function of metallicity and environment, and ultimately planet formation as a function of environment.

High spectral resolution (R~100,000) and high S/N (>100) spectro-astrometry will allow us to probe the inner regions of disks up to a few kpc from the Sun. Additional high spatial (AO-assisted) resolution and IFU capabilities (at nearly the diffraction limit) in the near-IR will be critical to study more distant systems within the Galaxy and the MCs.

This science case would also greatly benefit of a spectro-polarimetric mode to trace the magnetic field of the star, which likely plays a crucial role in the evolution of the circumstellar disk, in the jet launching mechanism, in the accretion of protostars and, ultimately, in the formation of planets.

*White Dwarfs.*

About 20,000 White Dwarfs (WDs) have been identified in SDSS (Kleinman et al. 2013). Over the next decade southern sky surveys such as VST, VISTA and Skymapper will identify even more candidate WDs at V<21, which will be confirmed by the next generation multi-object spectrographs. Gaia will also identify ~$10^5$ WDs and provide crucial parallaxes and proper motions.

These surveys will provide targets crucial for understanding a range of outstanding problems in WD science and where WDs probe wider topics in modern astrophysics.

<u>Double degenerate WDs as SNIa progenitors.</u> Despite their particular importance to cosmology and galaxy evolution, the likely progenitors of SNIa are poorly understood. Surveys of bright, nearby WDs have failed to reveal any candidate double degenerate binaries with sufficient mass that will merge on a Hubble time, yet recent studies of SNIa in nearby galaxies have provided support for this progenitor channel. Radial velocity studies of a much larger sample of candidate binary WDs revealed by new sky surveys and Gaia will be made possible by E-ELT HIRES medium and high resolution spectroscopy, potentially revealing, at last, the first genuine supernovae (SN) Ia progenitors and the contribution of this channel to the cosmological SNIa rate.

The empirical initial-final mass relation. The relation between the mass of a WD and its progenitor star remains relatively poorly constrained, particularly at the high mass end and for solar-type and lower mass stars. WDs in open clusters and moving groups of known age are best used to constrain the relation, but only those few in the nearest clusters are bright enough for 8m telescopes. E-ELT HIRES will enable spectroscopy of WDs in clusters with a wide range of ages and turn-off masses, providing temperatures, masses and cooling ages through modelling of the higher order lines in the Balmer series, and radial velocities to compliment Gaia proper motions to confirm cluster membership.

The mass-radius relation through eclipsing binaries. Detached, eclipsing binaries containing a WD and a low mass star or sub-stellar object, or even two WDs, and those in Sirius-like systems, provide model-independent examinations of the mass-radius relation for degenerate WDs. Currently a small number of suitable systems are known. New targets provided by southern sky surveys, together with WD parameters and radial velocities provided by the E-ELT HIRES will enable an examination of this fundamental relation across the entire range of WD masses, core and atmospheric compositions. HIRES will also allow detailed study of the cool companions in eclipsing systems, including examining the mass-radius relation for ultra-cool and brown dwarfs.

The main WD science requirements are as follows. (**a**) Blue spectroscopy to 380 nm, covering the Balmer series to high order lines, and the Ca K line. (**b**) Red spectroscopy for H$\alpha$ for radial velocities, and CaII 860.0 triplet as indicator of the presence of a gas disk. (**c**) Near-IR spectroscopy to study cool companions and reveal the presence of dust disks. (**d**) Medium resolution (~10,000) for white dwarf parameters (temperature, gravity, mass and cooling age) through Balmer line modeling, and for some radial velocity information. (**e**) High resolution (R~50,000) for radial velocities and to investigate metal line abundances (the latter is in common with the "debris" case in sec.2.4).

## 3.4. Stellar populations in the Local Group and beyond

The detailed study of stellar structure and evolution in different environments is of paramount importance to reconstruct the star formation and chemical enrichment history of galaxies and more generally, to understand galaxy and chemical evolution and properly quantifying impact and feedback. In this respect, the systematic study of Pop III stars and the oldest, unevolved Pop II stars is especially crucial for stellar archaeology and Local cosmology.

Also, a fundamental question is to which extent the stars of given age in a given stellar system show a real spread in chemical abundances, and to which extent true metallicity-age relations do exist. In order to establish such relations, and determine the physical scatter around them, relatively large samples of stars, observed at high spectral resolution and S/N and at different loci in the Galaxy and in the Local Group are needed.

Another question is to which extent stars formed in the same environment show different abundance patterns that could reflect limited mixing or differently polluted gas in the star-forming cloud. The accurate abundance signatures of stars associated with star-forming regions as well as in star clusters and in the galactic field could possibly disclose some particularities in terms of SN enrichment, mixing and dust cleansing of different star-forming events. Discoveries of such real spread in abundance patterns could be used for a more precise association of stars and groups of stars to different birth places. Indirectly, valuable constraints concerning the variation of the element yields of SN of various types may also result from this.

*Pristine chemical abundances and self-enrichment.*

A fundamental information in the study of stellar populations is the initial composition of the gas from which a given stellar system formed.

The extremely metal-poor (EMP) stars in our Galaxy are probably the most ancient fossil records of the chemical composition of the interstellar medium in the early times of the formation of the Milky Way and thus, indirectly, also on the pre-Galactic phases and on the stars that synthesized the first metals. The mode of formation of Pop III stars is expected to be different from that of "normal" Pop II stars. Masses and output of ionizing photons of Pop III stars can be inferred from the observed elemental ratios when compared with those expected from the explosion of stars formed from metal-free gas (Heger & Woosley 2010). One of the most crucial questions to be answered is the presence of Pop III low mass stars. Some theories predict that Pop III stars are necessarily very massive and that there is a critical metallicity above which the star formation switches to Pop II mode (see Bromm & Larson 2004, for a review), given that all the most metal poor stars discovered in the last decade are also C-rich (Christlieb et al. 2002, 2004; Bessell et al. 2004); Frebel et al. 2005, 2006; Aoki et al. 2006). However, the recent discovery of a very metal poor star (Caffau et al. 2011, 2012) with "normal" C and N has challenged these theories, showing that the formation of low mass stars with a primordial chemical composition could exist. Several surveys for metal poor stars are currently ongoing or planned. A few thousands EMP stars with [Fe/H]<−3.0 and ~100 with [Fe/H]<−5.0 are expected to be found by planned surveys but can only be verified and studied in detail with the sensitivity of E-ELT HIRES. While the H&K CaII lines can be also measured at R~10-20,000, essentially all chemical information is contained in weak spectral lines, including crucial species like FeII, C, N, O, S, Fe-peak and neutron capture elements, that require both high S/N from the E-ELT light-collecting power and high spectral resolving power (R>50,000).

Also the absence of significant star-to-star scatter in EMP halo stars (e.g. Cayrel et al. 2004, Caffau et al. 2011, 2012), given that these stars likely boast in the mean only one single progenitor, would imply a robust nucleosynthesis mechanism and/or a narrow mass range of (massive) zero-metallicity progenitors. Measurements of EMP stars in other halos are therefore of paramount importance to probe such a primordial nucleosynthesis.

Also a better understanding of Li production and destruction across the Galaxy and beyond, that is in systems having different chemical histories than the Galactic disk, like the Galactic halo and bulge, Sagittarius and the Magellanic Clouds has profound implications in our understanding of Big Bang nucleosynthesys and stellar physics. It is especially important to determine $^7$Li abundances in metal poor stars in different stellar systems and verifying the universality of the Spite Plateau, as well as to properly quantify $^7$Li production in AGB stars (de Laverny et al. 2006), in stellar systems spanning a wide range of [Fe/H]. Another important issue is verifying whether also $^6$Li show a Plateau (Lind et al., 2013).

The oldest, unevolved Main Sequence (MS) stars are the best witnesses of the pristine nucleosynthesis, after the primordial metal enrichment from Big Bang and Pop III stars. High resolution spectroscopy of these faint stars is out of reach with the current 8-10m class telescopes and an E-ELT is crucial to initiate their quantitative chemical screening across the Galaxy and out to Sagittarius and the Magellanic Clouds.

The detailed chemical composition of solar-type stars and cooler dwarfs will also be crucial to probe and quantify possible gradients (of each fundamental element or group of elements) across the various sub-components (halo, disk and bulge) of the Milky Way as well of its closest neighbours, leading to new and important constraints to mixing processes of gas and stars in their history, galaxy mergers, and nucleosynthesis in general. They will be very important complements to contemporary studies, which use evolved, bright giants or young early-type stars, HII regions, open clusters and planetary nebulae.

*Extra-galactic star clusters.*

Star clusters can be observed individually with E-ELT HIRES at distances of tens of Mpc, bringing a representative sample of galaxy types within reach of detailed stellar population studies. Clusters provide discrete sampling points in space and time, and trace environments ranging from starbursts and galactic disks to ancient spheroidal stellar populations. With internal velocity broadenings of only ~5 km/s, high-resolution studies of cluster integrated light reveal far more information than those of galaxies. R>50,000 and full spectral coverage will allow a complete chemical tagging of the cluster population and the measurement of its dynamical mass down to ~$10^4$ $M_\odot$, thus also constraining the stellar IMF of its host galaxy.

With the current generation of 8-10 m class telescopes, mostly radial velocities for hundreds of ancient globulars around giant ellipticals (Schuberth et al. 2010, Strader et al. 2011, Woodley et al. 2010) have been obtained. With E-ELT HIRES it will be possible to measure their detailed chemical abundances and dynamical masses and correlate these with the kinematic data, also providing crucial information on sub-structures (streams, shells, etc.), thus tracing possible accretion and/or merging events (e.g. Romanowsky et al. 2012). In Virgo and in Fornax galaxy clusters (Tamura et al. 2006a, 2006b), only the brightest GCs can be measured even with E-ELT HIRES, but in the nearest large elliptical, NGC 5128, all the clusters down to masses of ~$10^4$ $M_\odot$ can be observed. Similarly, young massive (from several $10^4$ $M_\odot$ to about $10^6$ $M_\odot$) star clusters can be probed out to distances of tens Mpc, tracing stellar populations in the disks of spiral galaxies and other star-forming environments, such as mergers and starbursts (e.g. from M83 to NGC4038/39, see Larsen et al. 2006).

The science cases in this section require full optical/near IR spectral coverage for complete line diagnostics and characterization of the stellar system properties. In particular, on the blue side, a wavelength coverage down to 3700-4000 Å would be desirable to measure H,K Ca (EMP science) , UII, CN; extension to even shorter wavelengths, 3100-3700 Å, would be useful to measure Balmer jump, NH, UV OH down to Be.

High resolution (R~100,000) and high S/N (>50) are required for a complete abundance screening of all the critical chemical elements and for precise stellar kinematics. Medium resolution (R~10-20,000) and high S/N (>50) are required for analysis of broad lines and for metallicity determination.

Some multiplexing capabilities as well as high spatial (AO) resolution (to resolve stars in crowded stellar fields like the centers of star clusters and nearby galaxies) are also desirable.

An E-ELT aperture is needed to get enough sensitivity for quantitative spectroscopy of (**i**) old Main sequence stars in the Galaxy and in the MCs, as well as of (**ii**) giant stars in the Local Group and beyond (Evans et al. 2011), thus probing stellar archaeology in a uniquely wide space of stellar and environment parameters and with adequate accuracy.

**Table.2.** Summary of science requirements for the main **stellar** science cases
(**E**=essential; **D**=desirable)

| Science case | | Spectral resolution ($\lambda/\Delta\lambda$) | Wavel. range ($\mu$m) | Wavel. accuracy (km $s^{-1}$) | Stability | Multi-plex | Backgr. subtr./ Flux cal. | AO / IFU | Polarim. |
|---|---|---|---|---|---|---|---|---|---|
| Stellar atmospheres | E | 100,000 | 0.37-2.4 | 0.3 | not critical | none | 0.5% | no | no |
| | D | 150,000 | 0.31-2.4 | 0.1 | not critical | AO/IFU | 0.5% | no | yes |
| Proto-planetary disks & proto-stellar jets | E | 100,000 | 0.38-2.5 | 1 | not critical | none | not crit. | AO | no |
| | D | 150,000 | 0.33-2.5 | 0.5 | not critical | none | <1% | AO+ IFU | yes |
| Stellar populations | E | 30,000 | 0.48-2.4 | 1 | not critical | 10 | not crit. | no | no |
| | D | 100,000 | 0.31-2.4 | 0.5 | not critical | 20 | not crit. | AO+ MOS[1] | yes |
| Extragalactic star clusters | E | 15,000 | 0.4-2.4 | 1 | not critical | none | <1% | no | no |
| | D | 30,000 | 0.37-2.4 | 0.5 | not critical | 10 | <1% | AO+ IFU/MOS | no |

[1] Dense stellar fields.

# 4. The Evolution of Galaxies and Cosmic Structures

## 4.1. Introduction

High-resolution spectroscopy is playing a crucial role in informing our understanding of the baryonic life cycle of galaxies. The investigation of Lyα absorption systems (Lyα forest, Lyman-Limit and Damped Lyα Absorbers (DLAs)), along the line of sight to quasars and Gamma Ray Bursts (GRBs) is one of the main tools of observational extragalactic astronomy and has delivered some of the seminal results in astrophysics. High-resolution absorption spectroscopy has enabled astronomers to probe the spatial distribution, the kinematics and the physical properties of baryons inside and outside of galaxies from the local universe out to the most distant quasars and GRBs discovered so far.

The sensitivity of this method owed to absorption spectra of intrinsically very luminous objects allows us to probe diffuse atomic hydrogen with column densities so low that they will be undetectable even with the most ambitious version of SKA. The wealth of transitions of the associated metal absorption by numerous atomic species, detected in absorption systems along the line of sight of high-z quasars and GRBs, have provided a detailed description of the chemical enrichment of the IGM and in the ISM of galaxies spanning a wide metallicity range ($3\mathrm{x}10^{-4}<Z/Z_\odot<1$), and with an accuracy that can again not be achieved with any other method. The observed enrichment patterns have provided detailed information on the generations of stars responsible for the production of the observed metals and have in this way provided tight constraints on the formation history of the stars responsible for producing them. The identification of the chemical fingerprints of the first generation of stars formed in the Universe with absorption line studies will be one of the main challenges in astrophysics in the next decade.

High-resolution spectroscopy has established the modern paradigm for the origin of the Lyα forest which we now know to probe mainly moderate amplitude density fluctuations in the large scale distribution of the photoionized Intergalactic Medium which can be expected (as demonstrated by detailed numerical simulations) to trace the underlying (dark) matter distribution very well. High-resolution spectroscopy has thereby established that the bulk of the IGM in the Universe is highly ionized and that the reionization of hydrogen was already largely complete by z~5-6. Determining how and when re-ionization proceeded and how the "dark ages" ended is one of the hottest topics in astrophysics, and requires high-resolution and high S/N spectroscopic observations of quasars and GRBs at z~7 and beyond.

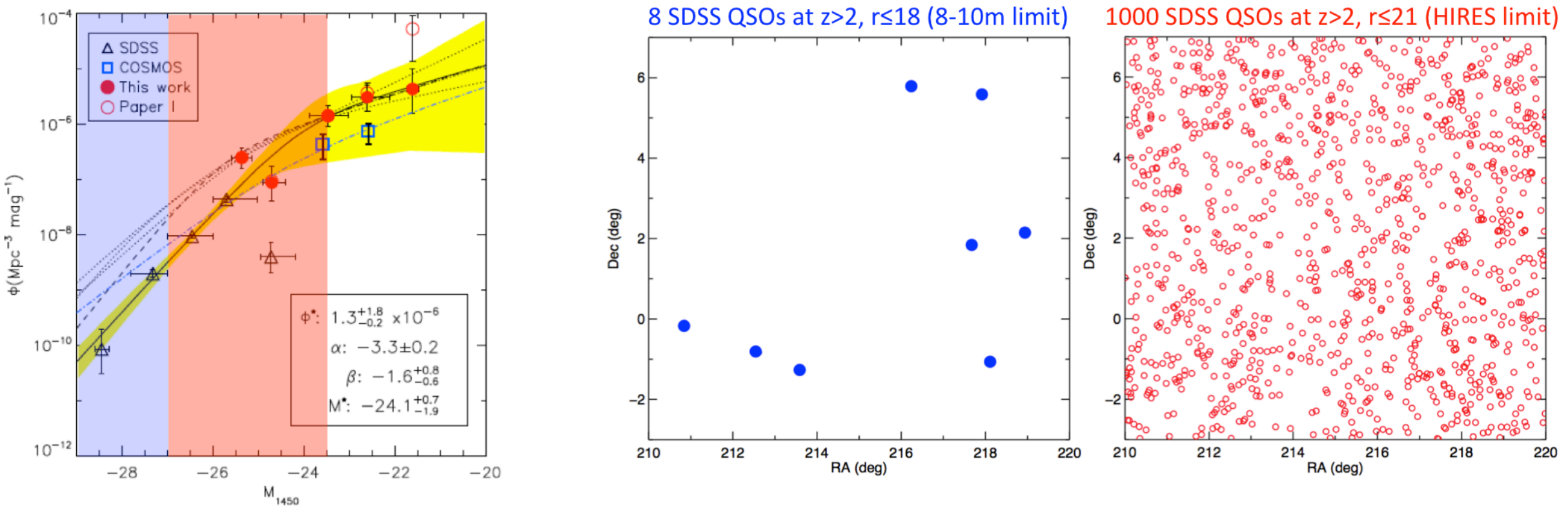

Fig.6. *Left:* Luminosity function of quasars at z~4. The blue shaded region shows the limit of high-resolution spectrographs at 8-10m telescopes, while the red shaded region shows the sensitivity extension enabled by HIRES at the E-ELT. Right: distribution and number of SDSS quasars at z>2 within reach of high-resolution spectrometers at 8-10m telescopes and of HIRES at the E-ELT, in an area of 10x10 deg$^2$.

High-resolution spectroscopy with 8-10m telescopes has reached the "photon-starving" regime in this field, and further progress desperately needs larger telescopes. By greatly expanding the number of observable quasars and GRBs, especially at high redshift, and by accessing much shallower and weaker spectral features, HIRES on the E-ELT will enable major discoveries that are not possible with current telescopes. A glance on the orders of magnitude sample "extension" enabled by HIRES is shown in the left panel of Fig.6, which shows the luminosity function of quasars at z~4. The blue shaded region shows the effective limit with current high-resolution spectrographs at 8-10m class telescopes, while the red shaded region illustrates the sensitivity extension enabled by HIRES at the E-ELT. It will be possible to reach quasars below L* and, given the steep luminosity function, this also translates into an increase by two orders of magnitude in the number of QSOs (and therefore e.g. DLAs) accessible with the E-ELT compared to current facilities. This is illustrated in the right panel of Fig.6, which shows the distribution and number of quasars at z>2 in a field of 100 deg$^2$ accessible to high resolution spectroscopy at 8-10m class telescopes (r<18) compared to what will be accessible to HIRES at the E-ELT (r<21).

HIRES will significantly push back the redshift barrier for studies of the IGM and ISM. As an example, Fig.8 shows the comparison between the best spectrum obtained of GRB 090423 at z=8.2 at the VLT and a simulated spectrum spectrum that could have been obtained if HIRES would have been operative. Clearly, HIRES would have enabled us to reveal the chemical enrichment and the ionisation state of the IGM pattern at this early time in the history of the Universe.

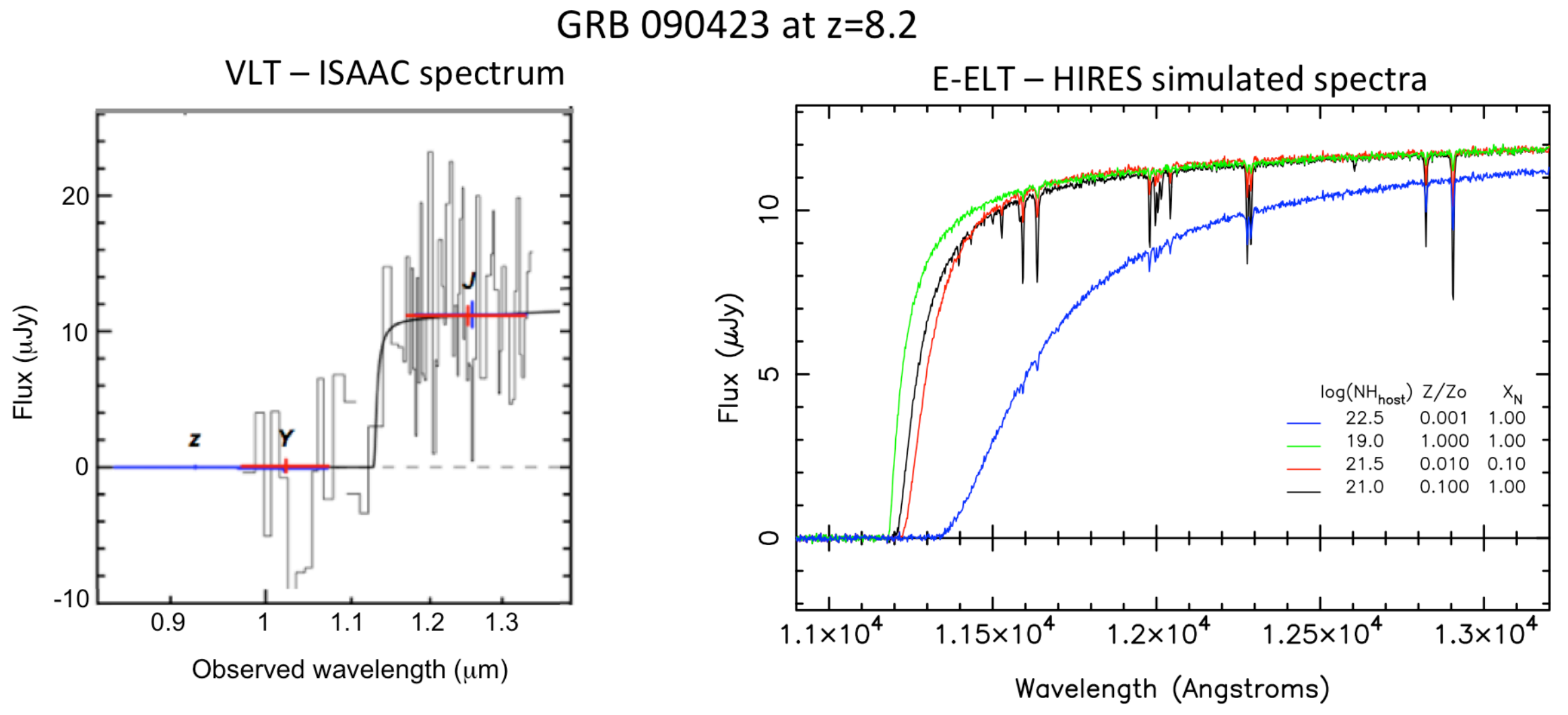

Fig.7. *Left:* Spectrum of GRB 090423 at z=8.2 taken with ISAAC at the VLT (Tanvir et al. 2009). *Right:* simulated HIRES spectra of the same GRB with various possible values of the colum of gas in the host galaxy, metallicity and IGM neutral fraction, which can easily be discriminated with such high S/N spectra. The metal lines represent typical species found in GRB spectra, and show that even very low metallicities can be recovered.

Even more interesting, HIRES make it possible to fully exploit Super-Luminous supernovae (SLSNe) for studies of the IGM and the ISM in distant galaxies. Unlike normal SN II or Ia, SLSNe are bright in the UV ($M_V \sim -22$ to $-23$ mags), and unlike GRBs, SLSNe are bright for a comparatively long time. They can therefore be identified and followed using traditional means for several months. This has the big advantage that they do not need a γ-ray and X-ray satellite for a fast identification and precise localization.

In the following, we illustrate key science cases for HIRES in the field of extragalactic astronomy. We will primarily focus on the discovery space opened in the areas of the early chemical enrichment and re-ionization of the Universe, but we will show that HIRES will also have a huge impact for the investigation of the evolution of massive galaxies and the co-evolution of supermassive black holes and their host galaxies.

### 4.2. Near-pristine Gas at High Redshifts

In the last few years it has come to be appreciated that the most metal-poor damped Lyman alpha systems, or DLAs, provide us with one of the best windows on early nucleosynthesis. DLAs are clouds of neutral gas most easily identified and studied at redshifts z = 2–4. At these cosmic epochs, ~2–3Gyr after the Big Bang, they span a range of metallicities, with a median value $Z_{DLA} \sim 1/20\ Z_{\odot}$, and with a low metallicity tail that extends at least to $Z \sim 1/3000\ Z_{\odot}$ or [Fe/H] ~ -3.5. It is these most metal-poor DLAs that are currently attracting a great deal of interest. They evidently arise in pockets of gas that have experienced minimal processing through stars, possibly at much earlier epochs than the redshifts at which we observe them, and may well be the high redshift counterparts of the gas from which the most metal-poor stars in the halo of the Milky Way condensed. Their chemical composition holds precious clues to the nature of the first stars that formed in the Universe. Measurements of element abundances in very metal-poor (VMP) DLAs complement and extend analogous studies in metal-poor stars of the Milky Way and its satellites (see sect.4.3). Indeed, VMP DLAs are arguably the best astrophysical environments for determining the relative abundances of the most common chemical elements, C, N, O, Mg, Si, S, Al, and Fe, synthesised in the first few cycles of star formation in the young Universe.

High resolution observations of VMP DLAs with UVES on the VLT and analogous instruments on other 8–10 m class telescopes have already clarified a number of unresolved issues left over from stellar work (e.g. Pettini & Cooke 2013). Recent highlights include:

1. Resolution of the ambiguity regarding the degree of alpha-element enhancement at the lowest metallicities, where different stellar features give discordant values of the O/Fe ratio. This ratio ultimately reflects the Initial Mass Function of the earliest stellar populations. The DLA measurements indicate only modest alpha enhancements, by factors of 2–3, consistent with a normal IMF.

2. The discovery of two DLAs exhibiting a marked excess in their C/Fe ratio, analogous to those found in the so-called 'Carbon-Enhanced Metal-Poor' (CEMP) stars. Unlike the stellar case, there is no ambiguity in the interpretation of such enhancements in DLAs, where they reflect the composition of the gas from which presumably some CEMP stars would later form (unlike the scenario whereby carbon-rich material was provided by a now extinct companion star).

While these initial results are exciting, they are also very frustrating. In a nutshell, this work has now reached the photon-starved regime. The observations highlighted above typically required the equivalent of one night of telescope time on the world's premier facilities (VLT, Keck, Magellan); consequently, it has taken several years to build-up even modest samples of VMP DLAs with well determined abundances. While large- scale imaging surveys of the sky continue to increase the number of known QSOs—and therefore damped Lyman alpha systems—the new additions to these catalogues generally involve QSOs which are too faint to be observed at high spectral resolution with current instrumentation.

Yet, there are burning questions waiting to be answered. In particular, with only two known examples of Carbon-Enhanced Metal-Poor DLAs it is impossible to draw general conclusions on the early generations of stars that seeded this gas. Also, the relative abundances of Fe-peak elements (i.e. Ti/Fe, Cr/Fe, Co/Fe, Ni/Fe, Zn/Fe) depend more sensitively on the masses, explosion energies and mass-cuts of the core-collapse supernovae that created them than the relative abundances of the more common alpha-capture elements. At present, however, absorption lines from these rarer species are too weak to be detected in individual VMP DLAs, and existing data are too few to give a definite result even with stacking techniques (e.g. Cooke & Pettini 2013).

Therefore, one of the most exciting prospects for E-ELT HIRES is the detection of elements synthesized by the first stars in the Universe. Massive, metal-free "Pop III" stars are expected to form in mini-halos at z > 15−20. Their radiation drives the first stages of reionization, and the heavy elements they produce allows gas to cool via metal-line radiation, enabling Pop II star formation. Individual Pop III stars are expected to be too faint to detect directly, even with JWST. Evidence for Pop III stars, however, may be found from their nucleosynthetic yields, as potentially observed in the abundance patterns of very metal-poor Damped Lyman-$\alpha$ systems.

E-ELT HIRES will enable spectroscopic surveys for Pop III enrichment directly in the reionization epoch. At higher redshifts, the elements produced by Pop III stars are less likely to have been diluted by those from subsequent generations of Pop II stars. Relative abundances have already been measured for a small number of low-ionization metal absorbers at z ~ 6, which have been found in the spectra of the ~10 z>6 quasars currently accessible at high resolution (Becker et al. 2012). These systems show similar abundance patterns as typical lower-redshift DLAs and non-carbon-enhanced metal-poor stars, suggesting that chemical enrichment at z ~ 6 is already dominated by conventional Pop II nucleosynthesis. A comprehensive search for metals from Pop III stars, therefore, must extent to fainter ($m_z \sim 21$), more numerous background quasars, and to higher redshifts. The known metal absorption lines at z~6 are typically narrow (b < 10 km/s), and weaker than those in lower-redshift DLAs. E-ELT HIRES will therefore be the ideal instrument with which to acquire a sufficiently large sample of z>6 absorbers (~20-50 systems) to determine the fraction of star formation in the reionization epoch that produced Pop III stars. Infrared capability beyond 1μm will be required to observe the primary lines of interest (e.g., O I 1302, C II 1334) at z > 6.7. An example of the capability of HIRES to reveal these tracers of the early chemical enrichment in the spectrum of a z~7 quasar is illustrated by the simulation of Fig.8, which also shows that this kind of science is out of reach for current facilities (e.g. X-shooter at the VLT).

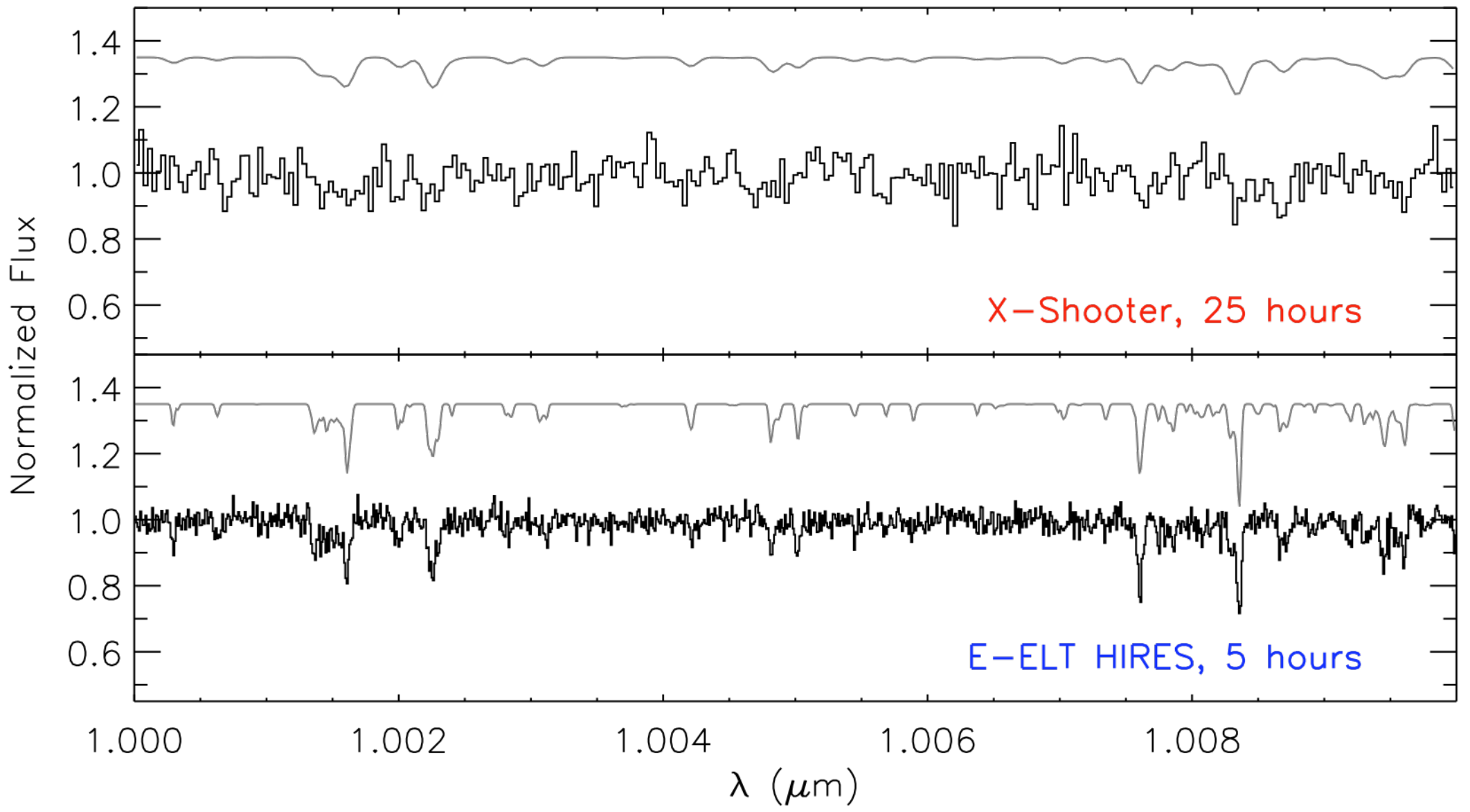

Fig.8. Simulated observations of an "O I forest" at z~7 using VLT+X-Shooter (top) and E-ELT+HIRES (bottom). The continuous grey line at the top of each panel shows the predicted absorption from O I, Si II, and C II generated from a cosmological simulation assuming an IGM neutral fraction of 10% and a metallicity of $10^{-3}$ $Z_{\odot}$. The signals have been smoothed by the instrumental resolutions. Signal-to-noise ratios have been calculated assuming a 25 (5) hour integration with X- Shooter (HIRES) on a z=7 source with $J_{AB}$ = 20.2, similar to the recently discovered quasar ULAS J1120+0641. No signal is detected in the X-Shooter spectrum. In contrast, numerous narrow lines are detected with HIRES.

All of these projects (and many other related lines of enquiry) will only become possible with a high resolution, R>50,000, spectrograph fed by a >30m UV-optical-NIR telescope, such as HIRES at the E-ELT. As mentioned, the increase by one order of magnitude in collecting area brings within reach of such an instrument two orders of magnitude more QSOs. In turn, this translates to a corresponding increase in the numbers of the intrinsically rare VMP DLAs, affording much improved statistics on all of the topics touched upon above.

Ideally the spectral resolution should be around 100,000 or higher in order to fully resolve the absorption systems. These can be as narrow as a few km/s, with broader systems often breaking into narrower subsystems. A resolution around 100,000 or higher would not only deblend individual components, but also enable measurements of the thermal and dynamical broadening of the absorbers, which are important tracers of ISM physics in these primordial systems. Yet, in the case of faint background sources a resolution higher than 100,000 may not provide enough signal-to-noise per resolution element in the continuum. In these cases rebinning to lower resolution (assuming detectors with very low read-out noise) is probably needed. Although a detailed tradeoff study is required to identify the optimal resolution for different classes of background sources, early estimates based on the faint tail of QSOs and GRBs at z>6 (known and expected from forthcoming facilities) suggest that R~50,000 is a good compromise between signal-to-noise in the continuum and the dilution of the absorption features by instrumental smoothing.

The spectral coverage is given by the need to probe different metals (to reconstruct the chemical enrichment pattern tracing the different possible stellar progenitors) in different ionization states. In particular, both low- and high-ionization lines are needed to correct abundances for ionization effects and to investigate the reionization process, as discussed in the next section. The primary low-ionization lines are Si II, O I, and C II with rest-wavelength 1260-1334 Å, thus coverage out to z=10 requires a wavelength range out to 1.5 μm. Fe II 1608 at z=10 requires a spectral coverage out to 1.8 μm, while observing the high-ionization doublet C IV 1548,1551 at z=10 implies $\lambda$ = 1.7 μm. It would also be highly desirable to observe the strong lines of singly ionized Fe and Mg, Fe II 2344-2600 Å and Mg II 2796-2803, as well as the neutral transitions Fe I (2484, 2523, 2967, 2984), Si I (2515), S I (1807) and Mg I (2026, 2853). These all require extension to 2.4 to be observed out to z=7.6.

## 4.2. Cosmic Reionization

The epoch of reionization was a transformational period in the early Universe. Within 300,00 years of the hot Big Bang, the Universe had expanded and cooled sufficiently to allow ions and free electrons to recombine, leaving the baryons almost completely neutral. Within one billion years, however, essentially all of the hydrogen in the Universe was once again ionized. The re-ionization of hydrogen is believed to have been caused by ultraviolet photons from the first stars and galaxies, most of which are too faint to be observed directly, even with JWST. Determining when and how reionization occurred therefore offers critical insight into both the history of baryons in the IGM and the formation of the first luminous objects.

The discovery of quasars out to z=6.4 in the Sloan Digital Sky Survey provided our first view of the IGM within one billion years of the Big Bang (e.g., Fan et al. 2001). Quasars at z~7 are now being identified (Mortlock et al. 2011), and infrared surveys such as VISTA and Euclid promise to uncover quasars out to z~8-10 and beyond. A high-resolution spectrograph on the E-ELT will deliver an unprecedented capability to probe the reionization epoch using these quasars as background sources, enabling detailed measurements of the ionization state, temperature, and chemical enrichment of the high-redshift IGM.

The large cross-section of neutral hydrogen to Lyman-$\alpha$ absorption means that a fully neutral IGM would produce an optical depth of $\sim 10^5$. The appearance of transmitted Lyman-$\alpha$ flux at z~6 in the spectra of high-redshift quasars, therefore, indicates that the IGM was largely neutral within one billion years of the Big Bang (Fan et al. 2006). Large sections of the z~6 Lyman-$\alpha$ forest appear to contain no flux, however, and while this is still consistent with a highly ionized IGM (e.g., Lidz et al. 2006, Becker et al. 2007), there remains the possibility that reionization has not fully completed at z~6 (McGreer et al. 2011), or that it ends at only slightly higher redshifts. Indeed, theoretical models suggest that reionization should be both extended and highly patchy (e.g., Furlanetto & Oh 2005). The integrated ionizing emissivity at lower redshifts, moreover, suggests that galaxies may struggle to completely reionize hydrogen by z~6 (e.g., Bolton et al. 2007).

Determining whether reionization truly extends to z~6 requires a substantially larger sample of high-quality z~6 quasar spectra. Samples of ~20 objects have been obtained using moderate-resolution spectrographs on 8-10m telescopes (Keck/ESI and VLT/X-Shooter; e.g., Fan et al. 2006, D'Odorico et al. 2011, 2013), ~10 of which have been observed at high-resolution with Keck/HIRES (Becker et al 2011b). Along with H I Ly-$\alpha$, these studies have targeted C IV, Si IV, C II, O I and Si II as proxies of the IGM ionization state. Only the spectra of the brightest objects, however, currently have enough sensitivity to place strong constraints on the neutral fraction. A robust characterization of the IGM ionization state at z~6-7 will require dozens of high-quality sight lines. In addition, the spectra must be observed at high-resolution in order to detect the narrow transmission peaks that are the first features to appear in the Lyman-$\alpha$ forest. Fig.9 shows that HIRES will be able to easily detect transmission peaks in quasars at z~6 with brightness down to m~21, undetectable with current facilities. Several dozens of sufficiently bright ($m_z < 21$) quasars are expected to be identified by southern sky surveys now in progress. The superb spectra obtained by E-ELT HIRES will determine whether reionization has fully completed by z~6, or whether variations in transmission between lines of sight indicate significant quantities of neutral gas.

The Thermal History of the IGM – As the IGM is reionized, photoionization heating will increase the gas temperature by ~ 10,000-20,000 K (e.g., Hui & Gnedin 1997). This heating increases the Jeans mass in the IGM, affecting the formation of low-mass galaxies. It also provides an additional avenue for probing both hydrogen and helium reionization. The changes in temperature alter the thermal broadening of Lyman-$\alpha$ absorption lines (Schaye et al. 2000), an effect that can be measured from high-resolution spectra. The small-scale features of the Lyman-$\alpha$ forest, therefore, can be used to trace heating during reionization, as well as the subsequent adiabatic cooling as the Universe expands.

The signature of extended helium reionization has recently been detected in the thermal evolution of the IGM in the redshift range $2 < z < 5$ (Becker et al. 2011a). Preliminary measurements have also been made of temperatures in the proximity zones of quasars at $z \sim 6$ (Bolton et al. 2010). Current measurements are limited, however, by the need to observe bright quasars in order to obtain sufficient sensitivity at high-resolution with 8-10m telescopes. Such sources are widely-spaced on the sky, by necessity, and are increasingly rare at higher redshifts. Observations with E-ELT HIRES will therefore dramatically improve our knowledge of the thermal state of the IGM in multiple ways. At $z < 5$, the ability to observe fainter targets will make it possible to use closely-spaced quasars groups to map the 3-dimensional topology of helium reionization, i.e., the growth and overlap of hot He III bubbles. At $z \sim 5 - 6$, HIRES observations will deliver precise temperature measurements based on the widths of rare Lyman-α transmission features. This will determine whether the IGM at $z \sim 6$ is still cooling following a late hydrogen reionization, or whether large-scale temperature fluctuations are present, as expected if reionization is both late and patchy. HIRES will deliver temperature information even at $z > 6$, using the Lyman-α absorption features in quasar proximity zones.

Such measurements will be a unique means of tracing reionization at higher redshifts where the Lyman-α forest itself is completely saturated, providing a powerful complement to redshifted 21-cm surveys.

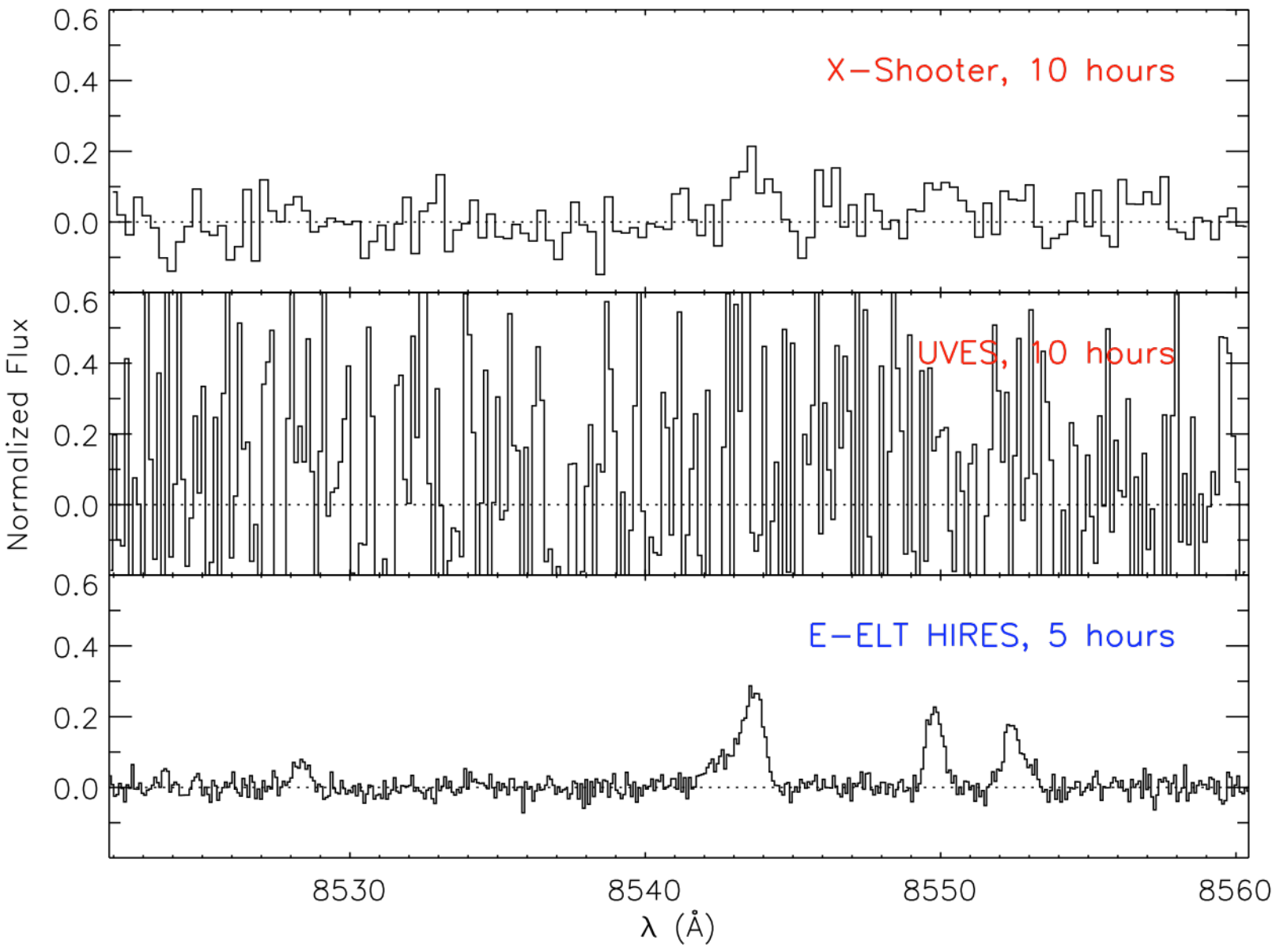

Fig.9. Simulated observations of a section of Lyman-α forest at z~6. Signal-to-noise ratios have been estimated for VLT+X-Shooter (top), VLT+UVES (middle), and E-ELT+HIRES (bottom) assuming a z~6 quasar with $m_z = 21$, typical of the faintest quasars found in SDSS. Marginal evidence of transmitted flux is present in the X-Shooter data, while UVES is completely dominated by read noise. E-ELT+HIRES, in contrast, robustly detects and resolves multiple narrow transmission peaks (note the weak feature at 8528 Å).

The instrument requirements are similar to those for the early chemical enrichment science case (previous section). A wavelength coverage up to 1.8μm is required to properly sample the transitions used to investigate re-ionization (C IV, C II, Si IV, Si II, O I) out to z~10. A wavelength coverage down to ~0.8 μm along with a spectral resolution of 50,000 is needed to detect the narrow transmission peaks shortward of Lyα.

### 4.3. Three-dimensional reconstruction of the Intergalactic Medium

The Intergalactic Medium (IGM) at high redshift, as probed by Lyα absorption systems, contains most (>90%) of the baryons in the Universe. Absorption line studies are therefore the method of choice to study the baryonic reservoir from which galaxies form. The galaxies emit ionizing photons and expel metals and energy and momentum through galactic winds powered by star formation and AGN, which in turn determine the physical state of the gas in the IGM. This interplay between galaxies and the IGM is central to an understanding of galaxy formation. Most of the interaction between galaxies and the IGM happens on scales of the order of 1 Mpc or less, or about ≤2 arcmin at z~2.5 and this region has been aptly termed the circum-galactic medium. At larger scales, above the Jeans-length of warm ($\sim 10^4$ K) photo-ionized gas, the IGM traces the underlying matter distribution in the linear regime.

By recording absorption systems on line of sights separated by a few arcminutes, it is thus possible to study the large-scale distribution of matter and to correlate the position of the galaxies with the density peaks marking the intersections of the cosmic web. In this way it will be possible to connect the topological structure of the cosmic web as characterized by statistical techniques (e.g. the PDF, the Euler characteristics, the skeleton, see e.g. Caucci et al. 2008) to the physical properties of the galaxies that form from it.

Faithful reconstruction of a representative part of the cosmic web by IGM "tomography", the dissection of the matter distribution by dense sampling with absorption spectra, requires sampling of a square degree or more with an angular separation of ~2 arcmin in order to (marginally) resolve the Jeans mass (Lee et al. 2013). As shown by Pichon et al. (2001) about 900 (randomly distributed) targets per square degree would be required for a reasonable reconstruction of the matter distribution.

Probing the transverse direction is not only important in order to remove systematic uncertainties due to the continuum fitting procedure, but offers a different way to recover the 3D distribution from the 1D cross-spectrum (Viel et al. 2002), due to the fact that 1D and 3D cross and auto flux power spectra are related to each other. These different approaches to probe the matter clustering and topology of the cosmic web will be eventually used also at low resolution and low signal-to-noise by the SDSS/BOSS collaboration (e.g. Slosar et al. 2009).

Historically, quasars have been the only suitably bright background sources to study the IGM at sufficiently high redshift ($z>\sim 2$) to have the Lyman alpha absorption above the atmospheric cutoff. However, quasars are not numerous enough to reach the density of 900 background sources per square degree required to obtain a proper reconstruction of the matter distribution. Lyman Break Galaxies down to a typical magnitude of R ~ 24, for which the E-ELT can obtain high-resolution spectra with a moderate amount of observing time, are the target of choice. The characteristic redshift will be $z\sim 2.5$ or larger, so that the far UV is not essential for this science case.

The cumulative density of LBGs in the redshift range $1.8<z<3.5$ reaches a value higher ~1 galaxy per arcmin$^2$ at a (Vega) magnitude R~24 (Steidel 2009). A field of view of 25 arcmin$^2$ ensures to have always (i.e. within 3 sigma of a random distribution) at least 10 targets per pointing, setting the desired level of multiplex. Sampling a representative volume of the Universe would require to cover an area of say 0.4 sq.deg of the sky with about 60 pointings to collect about 600 absorption spectra. This would enable an appropriate sampling of a few major filaments in a representative part of the IGM (e.g. Lee et al. 2013).

In terms of spectral resolution, in principle, R~5,000 is enough to recostruct the distribution of the IGM down to column densities $N_H \sim 10^{14}$ cm$^{-2}$. However, using LBGs as background sources to investigate the IGM poses some additional challenges compared to the use of QSOs. While the spectra of QSOs and GRBs are smooth and relatively featureless (the former having only very broad emission lines), the spectrum of LBGs at wavelengths shorter than Ly$\alpha$ is rich of stellar metal absorption lines (Rodriguez-Merino et al. 2005). These stellar lines will blend with the intervening IGM Ly$\alpha$ absorption features, hampering their detectability, if the spectral resolution is not high enough. Simulations are underway to identify the optimal spectral resolution needed to de-blend the IGM lines from the intrinsic LBG stellar lines. Preliminary results suggest that the spectral resolution should match the expected velocity dispersion of stellar clusters in LBGs (hence R~10,000) or even a factor of two higher (R~20,000), for a proper de-blending. Interestingly, with a resolution of R~10,000-15,000 the Jeans length along the line-of-sight is resolved and the resolution is thus a good match to the resolution across the line-of-sight.

We also note that such a project will provide hundreds of spectra with the quality of that of the seminal cB58 spectrum, but without the help of gravitational lensing.

It is also important to note that, while the 3D distribution of the IGM at high redshift will also be studied by other proposed facilities like the SKA, HIRES has the unique capability of mapping the *three-dimensional distribution of metals* within the cosmic web. Simultaneous detection of the metal absorption lines associated with Ly$\alpha$ systems is possible if the wavelength coverage extends to the near-IR. Such three-dimensional metallicity maps will provide the ultimate tool for an understanding of the metal enrichment process on large scales in the Universe. To achieve these goals a spectral resolution of R~10,000-15,000 is required over the wavelength range 0.4-1.3 μm, to enable the detection of the strong FeII and MgII absorption features out to at least $z\sim 3.6$.

## 4.5. Observing the different phases of massive galaxy formation and evolution

The formation and evolution of massive galaxies is a key open question in astrophysics and cosmology. In the present-day universe, most stars reside in massive early-type galaxies (ETGs) (e.g. Renzini 2006). Their stellar populations are the oldest amongst all galaxy types at all redshifts, they are often passively evolving and tend to inhabit the densest environments. ETGs are therefore excellent probes of the mass assembly in the hierarchical scenario of ΛCDM structure formation. Passively evolving ETGs should furthermore be reliable *cosmic chronometers* which allow to constrain the Hubble parameter *H(z)* (Moresco et al. 2012).

At $0<z<1$, the stellar mass function of ETGs shows a *downsizing* evolution difficult to reproduce with galaxy formation models (e.g. Pozzetti et al. 2010): most massive ETGs ($M_* > 10^{11}$ $M_\odot$) are already in place at $z \sim 0.7$, whereas the evolution is more pronounced for the lower mass ETGs, which increase substantially in number density from $z{\sim}1$ to $z = 0$, and faster in high-density environments (Bolzonella et al. 2010).

At $z > 1$, our knowledge of ETGs is still limited, but *bona fide* ETGs have been identified up to $z{\sim}3$ (e.g. Cimatti et al. 2004,2008; Kriek et al. 2006; Gobat et al. 2012). These high-z ETGs are characterized by old stars (~1-3 Gyr), short star formation timescales (~0.1-0.3 Gyr), low specific star formation rates (sSFR< $10^{-2}$ $Gyr^{-1}$), large stellar masses (up to > $10^{11}M_\odot$), and number densities growing rapidly from $z{\sim}3$ to $z{\sim}1$ (e.g. Domìnguez-Sanchez et al. 2011; Brammer et al. 2011). It is thereby not clear if the spheroidal morphology is a consequence of merging events, or is due to dynamical instabilities internal to the galaxies (e.g., Martig et al. 2009; Bournaud, Elmegreen & Martig 2009). The latter process may be particularly relevant given that the majority of ETGs are fast rotators (Emsellem et al. 2011). A puzzling property of ETGs at $z > 1$ is their small size, as small as $R_e < 1$ kpc, and correspondingly much higher internal mass density compared to present-day ETGs with the same mass (e.g. Cimatti, Nipoti & Cassata 2012 and references therein). The few available measurements of stellar velocity dispersions confirm that these systems are truly massive and dense (e.g. Cappellari et al. 2009; van Dokkum, Kriek & Franx 2009). Several models attempt to explain this size–mass evolution by including dissipationless (*dry*) major and minor merging, adiabatic expansion driven by stellar mass loss and/or strong feedback, and smooth stellar accretion, but a global and consistent picture has not yet been established. ETGs have also been identified in massive protoclusters at redshifts as high as $z \sim 2$ (e.g. Miley et al. 2006; Kurk et al. 2009; Gobat et al. 2011), but the role of the environment at $z>1$ is still uncertain. For instance, cluster cores at $1.3<z<2$ show clear signs of recent or on-going star formation, suggesting a reversal of the star formation - density relation.

The precursors of massive ETGs are therefore expected to be amongst high-redshift ($z{\sim}2$-4) starburst galaxies.

Their star-formation rates have to be sufficiently large (up to ~1000 $M_\odot\, yr^{-1}$) that they exhaust the gas reservoirs on short time scales in order to explain the required rapid, unsustainable growth at early epochs consistently with the observed properties at lower redshifts. AGN feedback may also play a key role in the rapid quenching of star formation. This should also be related to the observed black hole – galaxy mass relation of ETGs at $z \sim 0$.

Passive ETGs at $z>1$ are amongst the most difficult targets for spectroscopic studies. Since emission lines are absent or extremely weak, the physical, dynamical and evolutionary properties must be extracted from continuum and absorption line spectra. However, with ETGS being very red and very faint ($I_{AB}{\sim}24$-26, $K_{AB}{\sim}21$-23), continuum spectroscopy is only possible with the current largest telescopes for the very brightest objects of the population and with very long integration times (e.g. 30-60 hours in the optical or near-IR; Cimatti et al. 2008, Kriek et al. 2009). Moreover, some of the most useful spectral features (e.g. D4000 break, CaII H&K and Balmer lines, Lick indices) are redshifted into the near-IR at $z>1$.

The ESO VLT X-shooter spectrograph has demonstrated the power of high spectral resolution coupled with wide simultaneous wavelength coverage to investigate this class of objects (e.g. van de Sande 2011; Toft et al. 2012), although this work is still seriously limited by the "too small" collecting area of the VLT.

In order to study the physics of massive galaxies across all their evolutionary phases, from star-forming (+AGN) precursors at $z{\sim}2$-4 (where emission lines can also be studied) to old passive ETGs at $1< z < 2$, it is crucial to develop an instrument capable to extract all the information available in the spectra: stellar population properties through absorption lines (age, metallicity, star formation history), ionized and neutral gas characteristics through ISM emission and absorption lines (metallicity, ionization, density, temperature), dust extinction, star formation rate, AGN activity, gas kinematics (velocity dispersion, inflows, outflows, line profiles and asymmetries), stellar and AGN feedback effects (e.g. kinematic, radiative), mergers, scaling relations, role of cold flows, relationships between galaxies and gaseous halos. Deriving this rich information requires high-quality spectra (S/N>10 per resolution element) with moderately high resolution (R~10,000, both to trace dynamics and to maximise the sensitivity at λ>0.8μm, where OH airglow dominates) and simultaneous wide wavelength coverage (4000Å to 2.4μm). Fig. 10 illustrates the extraordinary amount of information that can be extracted with HIRES from high-z galaxies by accessing a broad range of diagnostics enabled by spectral coverage extending from the blue to the thermal near-IR. The same figure illustrates that the sensitivity of HIRES will allow to obtain such information for galaxies two orders of magnitude below M* at 1<z<3, and for statistically significant samples if some multiplexing is available.

Interestingly, HIRES will allow us to detect such features, especially the emission line diagnostics, also in the DLA galaxies that HIRES itself will identify in absorption in the foreground of QSOs or GRBs. This will make it for the first time possible to compare in detail the metallicity of the central, star forming-active, region of these galaxies, with their outskirts sampled by the absorption metal lines along the line of sight of the background QSO/GRB. This is currently achieved only for the brightest DLA galaxies (e.g. Krogager et al. 2013). HIRES will allow us to perform this kind of analysis for the whole population of typically much fainter DLA galaxies.

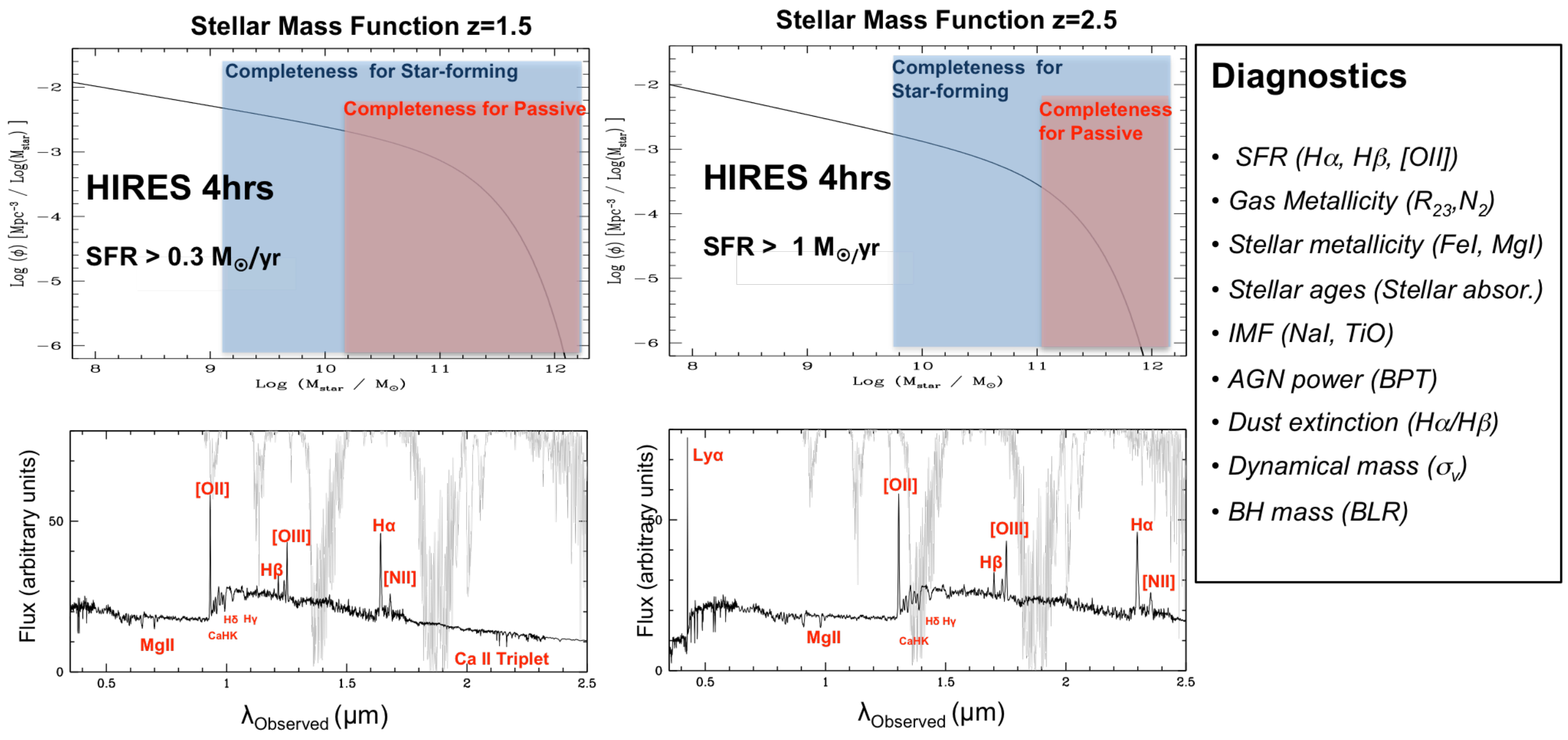

Fig. 10. Top left and top right panels show the Galaxy Stellar Mass Function at z=1.5 and z=2.5, respectively. The shaded regions highlight the completeness in stellar mass that can be reached by HIRES in just 4hrs of integration for star-forming galaxies (blue) and passively evolving galaxies (red) at those redshifts. With relatively short integrations it will be possible to fully characterise galaxies more than 2 orders of magnitude below M* in the redshift range 1<z<3, and building on the multiplex it will be possible to construct statistically significant samples of galaxies. In each panel it is also indicated the minimum star-formation rate than can be measured by HIRES in 4hrs at those redshifts, as low as 0.5-1 Msun/yr across the 1<z<3 range.
The bottom panels picture the typical spectrum of a star-forming galaxy at z=1.5 and z=2.5, showing the wealth of emission and absorption lines that are observed by HIRES simultaneously at those redshifts, enabling the measurements of key diagnostics highlighted in the legend on the right hand side. The grey curves show the atmospheric transmission.

Summarizing, the main requirements for the galaxy evolution cases with E-ELT HIRES are: simultaneous optical + near-IR coverage (0.4-2.4 μm; the optical coverage is essential to probe the rest-frame UV where several ISM and stellar absorption lines are present and to be sensitive to recent episodes of star formation in ETGs; $\lambda_{min}$~0.38 μm would be advisable for Lyα at z~2), moderately high resolution (R~10,000) to allow accurate measurements of spectral lines and their kinematical properties and to maximise the sensitivity at λ>0.8μm, where OH airglow dominates, MOS capability with N~10 multiplexing over a Field of View of 2-3 arcmin in order to study the dense regions of proto-clusters and the connections between galaxy halos and the IGM. Simultaneous wide wavelength coverage and high spectral resolution will make HIRES an instrument well suited for detailed studies of pre-selected galaxies in a way fully complementary to other potential E-ELT spectrographs.

## 4.6. Low mass Black Holes in galactic nuclei as tracers of the primordial seeds

One of the most striking achievements in the last decade has been the realization that supermassive black holes (BH; $M_{BH} \sim 10^5$ - $10^{10}$ $M_\odot$) are key actors in the formation and evolution of galaxies. This important fact is based on three seminal discoveries: the detection of supermassive black holes in the nuclei of nearby galaxies, the discovery that their masses are strongly correlated with the structural parameters of the host spheroids or galaxies in the case of ellipticals (Ferrarese & Merritt, 2000; Gebhardt et al., 2000; Marconi & Hunt, 2003) and that they were grown through efficient accretion during bright AGN phases (Marconi et al., 2004). In order to be established, $M_{BH}$-host galaxy correlations require a physical mechanism linking the pc-scale region surrounding the BH with the kpc-scale galaxy. Therefore, the physical mechanism triggering the gas accretion onto the SMBH is thought to be linked to global quantities characterizing the host galaxy (like star formation rate and stellar mass), possibly stabilized by a feedback mechanism. Such feedback from the actively accreting BH, i.e. the AGN, is also thought to contribute to the apparent anti-hierarchical growth of galaxies and supermassive black holes, the low fraction of baryons condensed into stars and the low number density of massive galaxies observed (Granato et al., 2004; Di Matteo et al., 2005; Hopkins et al., 2005, 2006; Menci et al., 2008). Understanding the resulting coevolution of supermassive black holes and their host galaxies is very important for understanding galaxy evolution.

The BH-galaxy coevolution paradigm is motivated by the observed correlations between BH mass and stellar velocity dispersion ($\sigma$), luminosity and mass of the host spheroid. These correlations are often taken to suggest that each galaxy hosts a supermassive black hole. However, this are far from being firmly established: most of the BH detections are in massive early type galaxies with $M_{BH} > 10^7$ $M_\odot$. Little is known for less massive galaxies which are predicted to host black holes with $M_{BH} < 10^7$ $M_\odot$, where only a handful of measurements have been obtained in the nearest galaxies with the Hubble Space Telescope and 8-10m ground-based telescopes with Adaptive Optics (AO) systems (see e.g. McConnell & Ma 2012 for a recent compilation of BH mass measurements).

It is currently believed that low mass galaxies undergo a quieter merger history than massive ones and, as a result, the *distribution of their black hole masses should retain an imprint of the original seed mass distribution*. Therefore, the fossil record of BH formation and early growth should be encoded in the demography of low mass black holes ($M_{BH} < 10^7$ $M_\odot$), but should be almost completely erased at larger masses (Volonteri et al. 2008). As a significant example, BH growth models predict significantly different $M_{BH}$ - $\sigma$ relations at $M_{BH} \sim 10^5$–$10^6$ $M_\odot$, for different initial mass functions of the seed black holes, that is for different seed formation mechanisms. Extending the census of black holes to $M_{BH} < 10^6$ $M_\odot$ would therefore allow us to study the BH-galaxy correlations at the low mass end, validate the coevolution scenario in low mass galaxies and constrain the formation mechanism of the seed BHs at this distance.

In order to populate the $M_{BH}$ - $\sigma$ relation at the low mass end, one needs to detect $10^5$–$10^6$ $M_\odot$ Black Holes at a distance of D~20 Mpc; this would require a survey of low mass galaxies in the Virgo cluster. A standard approach based on spatially resolved gas and/or stellar kinematics would require to spatially resolve the BH sphere of influence which, for a ~$10^5$ $M_\odot$ BH, in a galaxy with central stellar velocity dispersion of about 50 km/s located at distance D~20 Mpc, has a projeced diameter of 4 mas. The diffraction limited spatial resolution of a 39m E-ELT at 1.2μm, 2.2μm (e.g. at the location of Paβ, Brγ, H2) is 6 and 12 mas respectively, meaning that even the E-ELT will not be able to spatially resolve the sphere of influence of a $10^5$ $M_\odot$ Black Hole.

Spectroastrometry of gas emission lines is a method complementary to the classical techniques, that allows measurements at spatial scales smaller than the spatial resolution, by exploiting high spectral resolution. With spectroastrometry it is possible to measure gas rotational velocities on scales as small as 1/10 of the spatial resolution. The method has been tested and has been successfully used to measure BH masses (Gnerucci et al. 2010, 2012, 2013). At the heart of the method is the tracing of the spatial centroid of several, independent spectral channels spanning the emission line. By reaching even only 1/5 of the spatial resolution of the E-ELT, with spectroastrometry it will be possible to spatially resolve scales down to 1 and 2.5 mas at λ~1.2μm and λ~2.2μm, respectively. Spectroastrometry with the E-ELT will then therefore allow us to spatially resolve the sphere of influence of $10^5$ $M_\odot$ Black Holes at a distance of 20 Mpc. To adequately sample the rotation curve of a $10^5$ $M_\odot$ BH at D~20 Mpc, at least 10 spectral channels should be sampling the nuclear rotation curve (30-60km/s), therefore the spectral resolution must be (60-30)/10/2 = 3 km $s^{-1}$, corresponding to R~100, 000. To adequately map the centroid position of the central unresolved source an IFU sampling the diffraction limited E-ELT PSF is needed. Complete near-IR spectral coverage will allow the use of several gas emission lines, Paβ in J, [FeII] in H, and H2 and Brγ in K.

In summary, an IFU mode for HIRES working at the diffraction limit of the E-ELT and a spectral resolution of ~100,000 would be able to detect BHs with masses down to $10^5$ $M_{\odot}$ up to a distance of ~20 Mpc (i.e. Virgo).

**Table.3.** Summary of science requirements for the science cases related to **galaxy evolution** and **IGM** (**E**=essential; **D**=desirable)

| Science case | | Spectral resolution (λ/Δλ) | Wavel. range (μm) | Wavel. accuracy (λ/Δλ) | Stability | Multi-plex | Backgr. subtr. | AO / IFU | Polarim. |
|---|---|---|---|---|---|---|---|---|---|
| Near pristine gas & reionization | E | 50,000 | 0.6-1.8 | 50,000 | not critical | none | <1% | no | no |
| | D | 100,000 | 0.6-2.4 | 100,00 | not critical | 2[a] | <1% | no | no |
| 3D mapping of the IGM + metallicity | E | 5,000 | 0.4-1.3 | 5,000 | not critical | 5 | <1% | no | no |
| | D | 20,000 | 0.37-1.3 | 20,000 | not critical | 10 | <0.1% | no | no |
| Galaxy evolution | E | 10,000 | 0.4-2.4 | 10,000 | not critical | 5 | <1% | no | no |
| | D | 15,000 | 0.4-2.4 | 15,000 | not critical | 10 | <1% | no | no |
| Low mass Black Holes | E | 100,000 | 1-2.4 | not critical | not critical | none | not crit. | AO+ IFU | no |
| | D | 100,000 | 0.5-2.4 | not critical | not critical | none | not crit. | AO+ IFU | no |

[a] QSO pairs.

# 5. Fundamental Physics and Cosmology

## 5.1. Introduction

The physical cause of the observed late-time acceleration of the expansion of the Universe is currently unclear and it suggests that our canonical theories of gravitation and particle physics are incomplete, if not incorrect. Broadly speaking, the acceleration is generally attributed either to some exotic form of mass-energy with significantly negative pressure ("dark energy") or to a large-scale breakdown of the field equations of General Relativity describing gravity. In either case, the acceleration of the expansion points towards entirely new physics, and its discovery has sparked intense interest in mapping the expansion history of the Universe.

Observables that depend on the expansion history include distances and the linear growth of density perturbations, and so surveys of SNIa, weak lensing, redshift space distortions and baryon acoustic oscillations in the galaxy power spectrum are all considered to be excellent probes of the acceleration, which will be probed by ESA's Euclid mission, currently scheduled for launch in 2019. However, all of these probes are geometric in the sense that they seek to deduce the evolution of the expansion by mapping out our present-day past light-cone. None of these actually directly probe the global dynamics of the Friedman-Robertson-Walker metric.

After a quest of several decades, the recent LHC detection of a Higgs-like particle finally provides strong evidence in favor of the notion that fundamental scalar fields are part of Nature's building blocks. A pressing follow-up question is whether the associated field has a cosmological role, or whether there is another cosmological counterpart. A new generation of ground (ELTs, SKA) and space-based astronomical facilities (Euclid and others) will carry out precision consistency tests of our current paradigms in the near future and search for new physics beyond them.

The E-ELT can play a leading role in this endeavor. A range of instruments are expected to contribute (cf. MICADO's tests of gravity in the strong field regime and HARMONI's characterization of type Ia supernovas at redshifts $z > 2$), but it should be clear that the high resolution and stability of HIRES will make it an exceptional tool for fundamental physics. In this section we highlight four compelling science drivers in this area: tests of the stability of fundamental couplings, measurements of the redshift drift, measurements of the redshift dependence of the CMB temperature, and measurements of the primordial deuterium abundance. In addition to their intrinsic merit, they are also examples of key synergies with other facilities such as ALMA, Euclid, Planck, SKA and possible space-based gravitational wave detectors (such as DECIGO and BBO).

## 5.2. Fundamental constants: Mapping the dark universe

Nature is characterized by a set of physical laws and fundamental dimensionless couplings, which historically we have assumed to be spacetime-invariant. For the former this is a cornerstone of the scientific method (it's hard to imagine how one could do science at all if it were not the case), but for the latter it is only a simplifying assumption without further justification. These couplings determine the properties of atoms, cells, stars and the universe as a whole, so it's remarkable how little we know about them: we have no "theory of constants", that describes their role in physical theories or even which of them are really fundamental. Indeed, our current working definition of a fundamental constant is simply *any parameter whose value cannot be calculated within a given theory, but must be found experimentally.*

Fundamental couplings are known to *run* with energy, and in many extensions of the standard model they will also *run* in time and *ramble* in space (i.e. they will depend on the local environment). In particular, this will be the case in theories with additional spacetime dimensions, such as string theory. A detection of varying fundamental couplings would be revolutionary: it would automatically prove that the Einstein Equivalence Principle is violated (and therefore that gravity cannot be purely geometry), and that there is a fifth force of nature.

Moreover, even improved null results are important. The simple way to understand this is to note that the natural scale for cosmological evolution of one of these couplings (driven by a fundamental scalar field) is the Hubble time. We would therefore expect a drift rate of the order of $10^{-10}$ $yr^{-1}$. However, current local bounds coming from atomic clock comparison experiments (Rosenband et al. 2008), are already about 6 orders of magnitude stronger, and rule out otherwise dark energy models in the grossly simplified case of linear drifts with time. This is a sort of fine-tuning that might be related to that more commonly associated with dark energy.

In theories where a dynamical scalar field yields, say a variation of the fine-structure constant $\alpha \equiv 2\pi e^2/hc$, the other gauge and Yukawa couplings are also expected to vary. In particular, in Grand Unified Theories the variation of α is related to that of energy scale of Quantum Chromodynamics, where the nucleon masses necessarily vary when measured in an energy scale that is independent of QCD (such as the electron mass). It then follows that a variation of proton–electron mass ratio $\mu \equiv m_p/m_e$ is also expected, although the relative size of both variations will be model-dependent.

Recent astrophysical evidence from quasar absorption systems (further discussed below) suggests a parts-per-million spatial variation of the fine-structure constant α at redshifts 2 - 3, with no corresponding variation seen in μ. Although no known theoretical model can explain such a result without considerable fine-tuning, it should also be said that there is also no identified systematic effect that can explain it. The next generation of high-resolution spectrographs will be needed to clarify the issue.

The transition frequencies of the ubiquitous narrow metal absorption lines from quasar absorption systems are sensitive to α (e.g. Bahcall et al. 1967; Dzuba et al. 1999) and those of the relatively rare $H_2$ absorbers are sensitive to μ (e.g. Thompson 1975). In both cases, the different transitions have different sensitivities, so, observationally, one expects relative velocity shifts between transitions in an absorber, in a single spectrum, if the constants indeed vary. A relative variation in α or μ of 1 part per million (ppm) leads to velocity shifts of ~20 m $s^{-1}$ between typical combinations of transitions.

As already mentioned, the observational status of the field is intriguing. The early indications of a smaller average α in ~140 Keck/HIRES absorbers at $z$~0.5–3.5 (e.g. Webb et al. 1999; Murphy et al. 2003) have recently been combined with indications of a larger average α in ~150 VLT/UVES absorbers to yield ~ 4σ evidence for a dipole-like variation in α across the sky at the 10 ppm level (Webb et al. 2011; King et al. 2012). Several other constraints from higher-quality spectra of individual absorbers exist (e.g. Levshakov et al. 2007; Molaro et al. 2008), but none directly support or strongly conflict with the α dipole evidence.

For μ, just six $H_2$ absorbers have been studied with precision <10 ppm by various groups, with no current indication of variability at $z > 2$ where the Lyman and Werner $H_2$ transitions can be observed redwards of the atmospheric cut-off (e.g. King et al. 2008; Thompson et al. 2009; Malec et al. 2010; Wendt & Molaro 2012, Bagdonaite et al. 2013, Rahmani et al. 2013). At lower redshifts, and even in our Galaxy, more precise null constraints on μ-variation are available from radio- and millimeter-wave spectra of even rarer (i.e. colder, denser) clouds containing more complex molecules like ammonia and methanol (e.g. Flambaum & Kozlov 2007; Murphy et al. 2008; Kanekar 2011; Jansen et al. 2011; Levshakov et al. 2011; Ellingsen et al. 2012). Other techniques involving radio spectra typically constrain combinations of constants by comparing different types of transitions (e.g. electronic, hyperfine, rotational etc.; Murphy et al. 2001; Tzanavaris et al. 2007; Kanekar et al. 2012).

Fig. 11 shows the statistical precision on Δα/α achievable with future high-resolution spectrographs as a function of telescope diameter. The comparison is made against VLT/UVES for a sample of quasar absorption systems which yields an overall precision of 1 part per million (blue circle); recent results from UVES have similar statistical precision. If the same sample of absorbers were observed on a spectrograph+telescope with the same total efficiency, same resolution (R ≈ 50,000), for the same total observing time, the precision on Δα/α would be inversely proportional to telescope diameter (black solid line). Under the same assumptions, the expected precision for several future spectrographs/telescopes is shown as blue bars. Note that tripling the resolving power to $R \approx 150{,}000$ would not improve the precision by 3 times because, in most metal ion absorption profiles, many velocity components exist, with widths >1 km/s each, and are closely blended together; i.e. the sharpest features in most absorption profiles are already resolved at $R \approx 50{,}000$. Therefore, only a modest improvement is expected (red dashed line). It is essential to understand that avoiding *systematic* errors is critical to improving measurements of Δα/α in future. To achieve this, future instruments must enable much more accurate calibration than is currently possible with UVES.

The important design requirements for a HIRES instrument to constrain α- and μ-variation are specified in Table 5. Note that these are modest compared to the requirements for the redshift drift experiment discussed in the next section. The most difficult of them to accommodate will be the UV efficiency and very blue cut-off (~330 nm) required to detect a large number of $H_2$ transitions ($913 < \lambda_{rest} < 1100$ Å) towards the brightest quasars, i.e. those at lower redshifts, $z \sim 2$. This requirement may be alleviated if, for example, radio/millimeter-wave molecular (e.g. methanol) absorbers can be identified at redshifts $z \sim 1$–2.5. This requirement may also be less important if more bright quasars with $H_2$ absorbers are identified at redshifts $z \geq 3$ than the few currently known.

It's worth bearing in mind that ESPRESSO at the VLT should achieve many/most of the design requirements in Table 5 well before the E-ELT is built (Pepe et al. 2010). This implies that, apart from the obvious task of confirming at very high S/N the results emerging from ESPRESSO, the definition of an optimal fundamental couplings observational programme for HIRES will take into account ESPRESSO results. Broadly speaking, two opposite scenarios can be envisaged

- If ESPRESSO unambiguously confirms that α and/or μ vary, the role of HIRES will be to map out this variation over the widest possible range of redshifts and environments, as this information can be used to infer the dynamics of the underlying degrees of freedom (with uniform redshift coverage being particularly important).
- If no variations are found, then the tighter bounds obtained can still be used to constrain the physics of the dark sector of the universe; in this case the redshifts of interest (and the relative importance of α and μ measurements) may be different and depend on other cosmological results then available (including those from Euclid, which the E-ELT can nicely complement).

Naturally, the E-ELT still has the best chance of finding a new effect below the noise level of all other telescopes by virtue of its larger collecting area. HIRES also has a further advantage over ESPRESSO in its IR coverage, which may allow precise measurements well beyond $z$ = 4 (although relatively little work exists in this area, it has much unexplored potential). Finally, it must be emphasized that significantly improved measurements (both in terms of statistical and systematic errors) are also expected from ALMA, which will provide a crucial complement to the HIRES measurements.

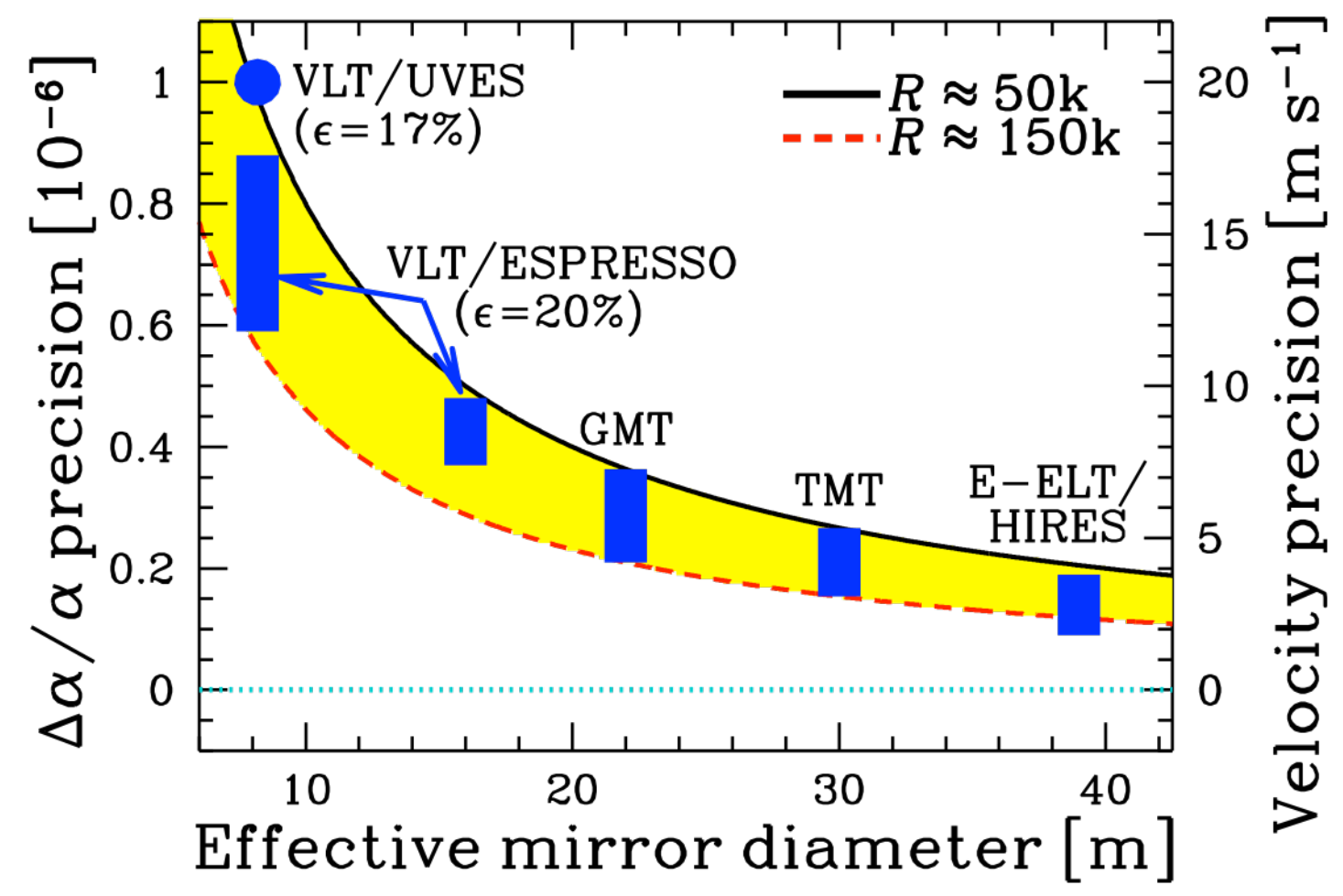

Fig.11. Expected statistical precision on variations in the fine-structure constant, Δα/α, achievable with future high-resolution spectrographs as a function of telescope diameter. The equivalent velocity precision is also shown (assuming a typical variety of metal-ion transitions). The length of the bars indicates the range of precision expected for different spectral resolutions available on those facilities. Two modes of operation for VLT/ESPRESSO are shown, its single-telescope mode (8-m effective diameter), with a range of resolving powers, and its anticipated four-telescope mode (16-m effective diameter), with $R$ up to ≈70,000.

If a dynamical scalar field is responsible for the recent acceleration of the universe one also expects it, as shown by Carroll (1998), to lead to potentially observable long-range forces and spacetime variations of the fundamental couplings. Conversely, measurements of these couplings (whether they are detections or upper bounds) can be used to constrain the dynamics of the dark sector of the universe. The recent acceleration of the universe suggests that any dynamics close to $z \sim 0$ will be heavily damped, we must probe the deep matter era regime, where the dynamics of the hypothetical scalar field is fastest. Fundamental couplings measurements are ideal for this task, and provide an ideal complement to lower-redshift probes such as Type Ia supernovas.

Amendola et al. (2012) have used Principal Component Analysis techniques to quantify the impact of ESPRESSO and HIRES on constraining dark energy through measurements of varying fundamental couplings, either by themselves or in combination with a SNAP-like supernova dataset (containing 3000 supernovas up to $z \sim 1.7$). Several possible scenarios were considered, some of which are shown in Table 4. The "Baseline", "Ideal" and "Control" cases correspond to different assumptions on the number and precision of the varying α measurements (in practice, to different amounts of telescope time).

In the case of HIRES, a reconstruction using α measurements can be more accurate than using supernovae data, its key advantage being the very large redshift lever arm. Since the two types of measurements probe different redshift ranges, combining them leads to a more complete picture of the evolution of the equation of state parameter. (Note also that a combination of JWST and HARMONI observations is expected to identify Type Ia supernovas up to $z\sim 4$.) These can realize the prospect of a detailed characterization of dark energy properties almost all the way up to redshift 4. Note that this analysis does not include redshift drift measurements, the inclusion of which can in principle further improve constraints.

Although the most obvious way to proceed is to combine the two datasets, we should also point out that they can be used separately to provide independent reconstructions. Comparing the two reconstructions will in fact provide a key consistency check for the assumptions on the coupling between the scalar field and electromagnetism and can be used to identify inconsistencies in the analysis, an example of which is discussed in Vielzeuf & Martins (2012). From the comparison one can obtain a measurement of the coupling parameter, which can be compared to those obtained from the CMB or from local Equivalence Principle tests. Interesting synergies also exist between these measurements and Euclid, which are currently being explored within Euclid working groups.

Last but not least, the importance of having measurements of several dimensionless couplings (such as α and μ) or combinations thereof cannot be overstressed, as these can provide unique tests of unification scenarios (which are complementary to, and in some sense more powerful than, those carried out at the LHC). These tests typically require measurements with accuracy better than parts-per-million, and with currently available data they can only reliably be done by using atomic clock measurements at $z = 0$, as discussed in Ferreira et al. (2012). HIRES will make similarly precise tests possible in the early universe.

| | Baseline | Ideal | Control |
|---|---|---|---|
| Supernovas | 2 | 2 | 2 |
| ESPRESSO ($\alpha$ only) | 0 | 1 | 0 |
| Supernovas + ESPRESSO | 2 | 3 | 2 |
| HIRES ($\alpha$ only) | 2 | 8 | 3 |
| Supernovas + HIRES | 4 | 10 | 5 |

Table 1: The table shows, for a dynamical dark energy model with a slow transition from a dust-like component at large redshifts to a cosmological constant-type behavior at low redshifts, how many PCA modes of the dark energy equation of state are well-determined (normalized uncertainty $\sigma < 0.3$) from observations of Type Ia supernovae (with a SNAP-like dataset) and/or α measurements from ESPRESSO and HIRES. See Amendola et al. (2012) for further details.

## 5.3. The Sandage test: Watching the Universe expand in real time

Sandage (1962) first discussed the possibility of observing that the redshifts of cosmologically distant objects drift slowly with time. If observed, their redshift drift-rate, *dz/dt*, would constitute evidence of the Hubble flow's deceleration or acceleration between redshift $z$ and today. Indeed, as emphasised by Liske et al. (2008), this observation would offer a direct, non-geometric, completely model-independent measurement of the Universe's expansion history and, to the best of our current knowledge, it would *uniquely* probe the global dynamics of the metric (Liske et al. 2008).

The observational challenge is that the drift-rate is expected to be extremely small: ~ 6 cm $s^{-1}$ $decade^{-1}$ at $z = 3$ (Liske et al. 2008). Nevertheless, Loeb (1998) proposed the Lyα forest of absorption lines seen towards background quasars as a promising target: the lines are numerous, ubiquitous, reasonably narrow (~10-30 km $s^{-1}$) and should be sufficiently immune to peculiar accelerations. HIRES should be the first instrument to meet this challenge. The theoretical interest in redshift drift measurements is rapidly growing. Darling (2012) has recently suggested HI 21-cm lines from intervening galaxies along quasar sightlines as another probe, with a measurement using the future Square Kilometer Array (SKA) radio-telescope over a timespan of 5-12 years. Yagi et al. (2012) have argued that gravitational wave observatories such as DECIGO and BBO can measure the drift at low redshift ($z\sim 0.5$) using gravitational wave detections of neutron star binaries, possibly on an even shorter timespan (although a full feasibility study of this method remains to be performed). Combining all would allow a direct measurement of the expansion history of the Universe, all the way from the low-redshift accelerated phase to deep into the matter era (beyond z=4) with no model-dependent assumptions beyond that of homogeneity and isotropy.

Liske et al. (2008) addressed the in-principle (i.e. photon-limited) feasibility of measuring the redshift drift of the Lyα forest over ~2 decades with a 40-m class optical telescope like the E-ELT. The main conclusion is that, given the number of known, bright (V~15-17 mag), high-redshift (z~2-4) quasars, a 39-m E-ELT with high efficiency (ε = 25 %) could detect the redshift drift of the Lyα forest with 4000 hr observing time over a period of ~23 yrs. At first, 4000 hr seems like a rather large amount of observing time, but it is important to take into account here, that the extremely high signal-to-noise ratio (S/N) spectra of ~25-50 quasars would offer other unique scientific opportunities in the process. It is further worth noting that (i) a simple search of the VLT/UVES archives reveal that more than 3000 hrs of quasar spectroscopy have been performed with UVES over the last decade; and (ii) once a first epoch of quasar observations is made, the redshift drift signal grows linearly with time. Given the E-ELT's superior light- gathering power, a highly stable, extremely well calibrated HIRES instrument will offer the exiting opportunity to start performing the Sandage test. One other important practical issue bearing on the feasibility, time-scale and observing time required for a Sandage test with HIRES is the number of bright quasars at z>2 observable with the E-ELT. Liske et al. (2008) emphasised that the discovery of just a few very bright (V~15.5-16.2 mag) quasars at z~2.7–3.2 could substantially reduce the observing time required and/or the duration of the test. Future southern all-sky surveys may well identify such objects: their colours are very similar to stars, so previous surveys may have missed them.

It should nevertheless be obvious that exploiting the full potential of this fundamental experiment will almost certainly not only require more than one generation of astronomers/astrophysicists but also telescopes. Excitingly with the introduction of modern laser technology into astronomical wavelength calibration the technology is ripe now to set the clock running.

The relevant design requirements are specified in Table 5. It is important to note the following:

*Wavelength calibration stability and accuracy.* To ensure that systematic effects in the wavelength calibration of each quasar exposure are well below the ~20 cm $s^{-1}$ signal in *dz/dt* at z~3 over a ~20 yr time-scale, the wavelength calibration *absolute* accuracy has to be ~2 cm $s^{-1}$. If the wavelength calibration is absolute, the spectrograph needs not be stable to that accuracy for longer than a typical quasar exposure. Laser frequency combs (LFCs) can provide the necessary calibration information per exposure (Murphy et al. 2007) and prototype systems have been demonstrated (e.g. Steinmetz et al. 2008; Wilken et al. 2010; Murphy et al. 2012; Wilken et al. 2012). They also offer the possibility of absolute calibration, though not without additional challenges (e.g. Murphy et al. 2012).

*Ultra-violet efficiency/cut-off.* Liske et al. (2008) emphasised that, if the specific scientific goal of the Sandage test is the best possible constraint on $\Omega_\Lambda$ *from global dynamics,* then observing the Lyα forest over a large redshift range is important. In particular, observations at z~1.9–2.2 (where *dz/dt* is expected to change sign) compared with *dz/dt* measurements at z ≥ 3.5 provide the strongest constraints on $\Omega_\Lambda$. Thus, high UV efficiency and a relatively blue cut-off (~350 nm) would be highly desirable. This requirement may be alleviated if an SKA measurement were performed at z~0.6 (Darling 2012), where *dz/dt* is smaller in magnitude but opposite in sign to that at *z*~3.

*Resolution.* Lyα forest lines have widths σ ≥ 7 km $s^{-1}$, so only a moderate resolving power of *R*~20,000 is needed to resolve them. However, to achieve a wavelength calibration precision of ~2 cm $s^{-1}$ per exposure, *R* ≥ 100,000 is required to provide a high enough density of calibration information (Murphy et al. 2007).

*Fibre scrambling.* With the simultaneous calibration approach (i.e. quasar & LFC light recorded next to each other on the CCD simultaneously), the quasar and LFC point-spread-functions (PSFs) *must* be essentially the same and the quasar PSF *cannot* depend significantly on the details of the acquisition, guiding or seeing variations. Fibre scrambling is therefore essential, and the efficiency with which angular movements are scrambled must be $\varepsilon_{scramble} > 2000$ to avoid significant miscalibration.

Redshift drift measurements by HIRES will be important to probe the redshift range $0.5 < z < 5$, thereby probing dynamical dark energy beyond the regime where it is dynamically important. Martinelli et al. (2012) recently quantified the gains of combining redshift drift and cosmic microwave background data (for which a Planck-like dataset was assumed). Fig. 12 summarizes these results. Combining the two probes significantly improves constraints on the geometrical cosmological parameters: CMB results on $w_0$, $H_0$ and $\Omega_m$ can be improved by factors of about 3, 3 and 2 respectively. These improvements are not due to a better sensitivity of the redshift drift test, but to different degeneracies. For example, the CMB is sensitive to the combination $\Omega_m H_0^2$, but the redshift drift probes the two parameters separately. Interestingly, the redshift drift can be a better probe of the *matter* density than of the dark energy one, if the appropriate redshift range is used (Liske et al. 2008). Doing the analysis with a phenomenological $w_0$ - $w_a$ parametrization of dark energy one finds that the constraints on these parameters are not significantly improved by redshift drift measurements to CMB data, while those on $H_0$ and $\Omega_m$ are improved as above. This means that the redshift drift is not able to break the existing degeneracy between $w_0$ and $w_a$. To break this degeneracy one would need to add low-redshift measurements of the redshift drift, which SKA may provide.

Vielzeuf & Martins (2012) have also shown that combining redshift drift and precision spectroscopy measurements of varying couplings (discussed above) gives HIRES a further advantage: it allows it to test classes of cosmological models that would otherwise be difficult to distinguish from ΛCDM (because they would be extremely similar at low redshifts). It will also allow it to check the consistency of reconstructions of the redshift dependence of the dark energy equation of state. Last but not least, there are also important synergies between redshift drift measurements and Euclid, which are presently being quantified within Euclid Consortium working groups.

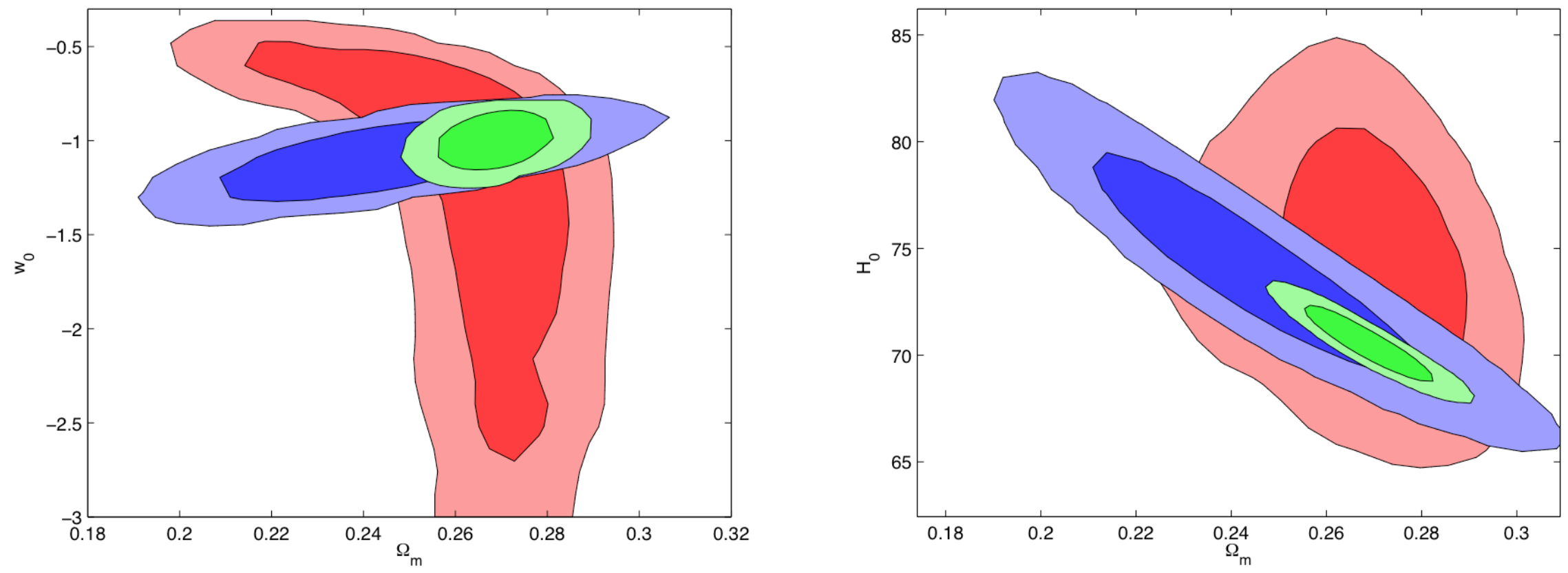

Fig.12. 2D constraints on the $w_0$-$\Omega_m$ and $H_0$-$\Omega_m$ planes using CMB (blue), redshift drift (red) and combining the two probes (green). Reproduced from Martinelli et al. (2012), where the precise assumptions are described.

### 5.4. The CMB temperature: Mapping the bright universe

When one finds direct evidence for new physics, it will only be believed once it is seen through multiple independent probes. It is therefore crucial to develop consistency tests; in other words, astrophysical observables whose behaviour will also be non-standard as a consequence of varying couplings or dynamical dark energy.

One of the most precise measurements in cosmology is the intensity spectrum of the cosmic microwave background (CMB) radiation: at $z = 0$ one finds a very precise black-body spectrum. However, this measurement tells us nothing about the behaviour of the CMB at non-zero redshift. The CMB temperature-redshift relation, $T(z) = T_0(1 + z)$, is a robust prediction of standard cosmology; it assumes adiabatic expansion and photon number conservation, but it is violated in many scenarios, including string theory inspired ones, as well as models where α varies. Phenomenologically one can parametrize deviations to this law by adding an extra parameter, say $T(z) = T_0(1 + z)^{1-\beta}$. The recent spectroscopic measurements at redshifts $z \sim 2$ - $3$ by Noterdaeme et al. (2011), combined with measurements of the SZ effect at resdshifts $z < 1$, yield the $\beta = -0.01 \pm 0.03$.

The distance duality relation, $d_L(z) = (1+z)^2\, d_A(z)$ is an equally robust prediction of standard cosmology; it assumes a metric theory of gravity and photon number conservation, but again is violated if photon number is not conserved. In fact, in many such models (including varying-α ones) the temperature-redshift relation and the distance duality relation are not independent: a direct relation exists between violations of to the two laws. This link allowed Avgoustidis et al. (2012) to use distance duality measurements to further constrain β, yielding $\beta = 0.004 \pm 0.016$, which is a 40% improvement on the direct constraint. Moreover, the size and redshift dependence of this violation *is also related to the evolution of* α.

The current percent-level constraints on β do not yet have a strong impact on theoretical models, Progress in the number and sensitivity of low-redshift *T* (z) measurements will come from using clusters in the Planck SZ catalogue and also from ALMA, but HIRES will be the key instrument for improving the high-redshift spectroscopic measurements and thus (due to the large redshift lever arm) ultimately provide a crucial consistency check for any claims of α or μ variations.

The CMB temperature at $z > 0$ can be measured spectroscopically by using the excitation of interstellar atomic (like CI) or molecular species (like CN or CO) that have transition energies in the sub-millimetre range and can be excited by CMB photons. When the relative population of the different energy levels are in radiative equilibrium with the CMB radiation, the excitation temperature of the species equals that of the black-body radiation at that redshift. As first shown by Bahcall & Wolf (1968), the detection of these species in diffuse gas, where collisional excitation is negligible, provides one of the best thermometers for determining the black-body temperature of the CMB in the distant Universe.

CO has been detected recently in diffuse gas at high redshift; as discussed in Noterdaeme et al. (2011) these provide a sample of *T* (z) measurements with uncertainties of order 1 K or slightly better, which correlate linearly with the spectral quality. These can be improved significantly by HIRES and hopefully new CO systems will be discovered in the near future. The specific requirements are similar to those for μ measurements: one needs high resolution (R > 100,000) and a good UV-blue throughput is also important.

It's also important to realize that these measurements are not necessarily costly in terms of telescope time, since in many such systems μ can also be measured. This is potentially important in case of detections of deviations from the standard behavior: as pointed out above, if a variation of μ is found one will also expect a breakdown of the standard temperature evolution (by an amount that is precisely calculable, though somewhat model-dependent) so looking for it in the very same system would be an ideal consistency test.

Avgoustidis et al. (2012) have described how several forthcoming facilities can lead to much stronger constraints. HIRES can deliver a sensitivity nearly an order of magnitude better than current probes, and a factor of three better than expected from ESPRESSO. Ongoing (currently unpublished) work suggests that this is sufficient to place non-trivial constraints on models leading to variations of fundamental couplings at the level of the α dipole.

## 5.5. The abundance of deuterium and constraints on non-standard physics

It is well known that the relative abundances of the light elements created in the first few minutes of cosmic history depend on the product of the present-day density of baryons in units of the critical density, $\Omega_{b,0}$, and $h^2$ (where $h$ is the present-day value of the Hubble parameter measured in units of 100 km $s^{-1}$ $Mpc^{-1}$). When analysed in conjunction with the power spectrum of temperature anisotropies of the cosmic background radiation, light element abundances can improve constraints on cosmological parameters, primarily the spectral index of primordial fluctuations (Pettini et al. 2008), and on the effective number of light fermion species, $N_{eff}$ (e.g. Simha & Steigman 2008; Nollett & Holder 2011).

Among the light elements created in Big Bang nucleosynthesis (BBN), deuterium is one of those that have attracted the attention of many recent works. In particular, high resolution spectroscopic observations of metal-poor hydrogen clouds at high redshift have delivered the most precise measurement yet of the abundance of deuterium produced in Big Bang Nucleosynthesis. Its agreement with the cosmic density of baryons, $\Omega_{b,0}$ , deduced from the temperature fluctuations of the CMB places limits on possible departures from the standard model of particle physics. For example, the number of neutrino families required to preserve consistency between $\Omega_{b,0}$ (BBN) and $\Omega_{b,0}$ (CMB) is limited to $N_\nu = 3.0 \pm 0.5$, placing into doubt the existence of so-called "dark radiation".

HIRES will facilitate several new precise measurements of the primordial abundance of deuterium, thanks to the much larger sample of accessible quasars (hence DLAs), especially at high redshift (see sect. 4.1). These will be needed to match the accuracy of the CMB observations that are being released by the Planck mission. Only by comparing determinations of $\Omega_{b,0}$ of comparable precision, but sampling different epochs in the expansion history of our Universe (i.e. the first few minutes with BBN and ~ 370000 years after the Big Bang with the CMB), will it be possible to place the most stringent constraint on non-standard physics (Hamann et al. 2011).

The most useful D I lines are those close to the limit of the Lyman series, because they are the weak, unsaturated ones,

which give the most precise estimate of the column density of D I. To access these lines in absorption systems at z~3, as in previous studies (Pettini & Cooke 2012), requires a wavelength coverage at least to 3700 Å. However, as illustrated by Pettini & Cooke (2012), many of the best Deuterium abundance measurements are obtained at even lower redshifts, down to z~2.6, indicating that a wavelength coverage extending to 3300 Å would be desirable.

**Table.5.** Summary of science requirements for **fundamental physics** and **cosmology** (**E**=essential; **D**=desirable)

| Science case | | Spectral resolution ($\lambda/\Delta\lambda$) | Wavel. range ($\mu$m) | Wavel. accuracy (m $s^{-1}$) | Stability (m $s^{-1}$) | Multi-plex | Backgr. subtr. | AO / IFU | Polarim. |
|---|---|---|---|---|---|---|---|---|---|
| **Fundamental constants & T(CMB)** | **E** | 80,000 | 0.37-0.67 | 2 (relative) | 2 $night^{-1}$ | none | not critical | no | no |
| | **D** | 100,000 | 0.33-0.8 | 1 (relative) | 1 $night^{-1}$ | none | desirable | no | no |
| **Deuterium abundance** | **E** | 50,000 | 0.37-0.7 | 50 | not critical | none | not crit. | no | no |
| | **D** | 100,000 | 0.33-1.0 | 50 | not critical | none | <1%[a] | no | no |
| **Sandage test** | **E** | 100,000 | 0.37-0.67 | 0.02 (absolute) | 0.02 $night^{-1}$ | none | not critical | no | no |
| | **D** | 150,000 | 0.33-0.8 | 0.01 (absolute) | 0.01 $night^{-1}$ | none | desirable | no | no |

[a] Faint quasars limit.

# 6. Summary of the science requirements

One of the main emerging requirements for many of the science cases is the need for a broad spectral coverage, extending from the blue to the K-band. Possibly, this large spectral coverage should be covered *simultaneously*, both because to cost of a E-ELT is so high (>1 G€) that any mean to save observing time should be pursued, and because the experience gained with X-Shooter has revealed that the inter-calibration of the spectra in different spectral ranges is much more difficult and inaccurate if taken at different times (both because of the variable observing conditions and because of intrinsic variability of astronomical targets), in many cases so bad to preclude the achievement of the science cases. At the blue extreme of the spectral range, for some of the science cases an extension to wavelengths shortward of 0.37 μm is desirable, but not essential. Hence, a coating of the mirrors optimized for the extreme blue is not an essential requirement for the science cases of HIRES. It is more important to note that for the bulk of science cases high sensitivity in the near-IR is crucial.

The bulk of science cases require a spectral resolution R~100,000. Most of the science cases do not require high calibration accuracy (wavelength calibration accuracy below 1 m $s^{-1}$) and most science cases need the wavelength calibration stability requirements to be met only within the observing night. It is interesting to note that the new focus on the exoplanet science case onto exo-atmospheres has shifted the stability requirement onto the detector and the flat-fielding, rather than wavelength.

The only science cases with stringent constraints on the absolute wavelength calibration accuracy is the direct measurement of the Universe acceleration (Sandage's test), which requires an absolute calibration of about 2 cm $s^{-1}$. For the exo-planets radial velocities this requirement is relaxed to 10 cm $s^{-1}$.

Some science cases, such as the three-dimensional mapping of the IGM, galaxy evolution and spectroscopy of resolved stellar populations/star clusters in external galaxies, would need to exploit the same broad spectral coverage at intermediate resolution (R~10,000-15,000), but with some multiplexing, of the order of 5-10 objects over a field of view of a few arcminutes.

The high-z science cases, dealing with faint target, require a good sky subtraction (at the level of <1%), which we suggest is best achieved through the nodding technique widely adopted in near-IR, and which is known to remove the background to better than 0.1%, and by possibly sampling the sky with more than one spatial element (to minimize the sky noise).

The bulk of the HIRES science cases do not require spatially resolved spectroscopic information, hence seeing limited performances are adequate in most cases. The only cases requiring (at R~100,000) diffraction limited performances, at least in the near-IR, are the investigation of protoplanetary disks and protostellar jets, as well as the measurement of low mass black holes in galactic nuclei. For these observations an IFU mode, sampling the E-ELT diffraction limit at near-IR wavelengths, would be highly desirable.

Some science cases, including the characterization of exoplanets, stellar photospheres and protostellar accretion disks and protostellar jets, would benefit by enabling HIRES with a spectropolarimetric mode.

It is beyond the scope of this White Paper to provide detailed technical solutions for HIRES, which will have to be addressed during "phase A". However, in the following Annex we outline a possible preliminary concept for HIRES, which is not meant to be a proposal, but only to illustrate that simple technical solutions can be envisaged to meet all scientific requirements emerged from the various science cases described in this White Paper.

# ANNEX

## A preliminary concept for HIRES

We show here that the various science requirements outlined in the previous sections, and summarized in Tables 1-5, can be met through a highly modular concept, where different independent modules enable different observing modes and give access to different wavebands. A modular concept has several advantages.

i. Versatility. It can be developed in sequential stages, starting with the essential and highest-priority modules, and later upgrading of the instrument with additional less time-critical modules, depending on budget and resources.
ii. It reduces the overall risks associated with the project. Technical problems associated with an individual module do not delay the delivery of the other independent modules and their implementation at the telescope. Similarly, shortfalls of funding or resources experienced by one of the consortium partners will only affect the respective module this partner is involved in, and does not prevent the other partners to keep to the delivery schedule of the other modules.

We propose here that such a modular and versatile system can be realized by a system of fibers that feed simultaneously (through dichroics) different, independent spectrometers optimized for different wavebands. Each spectrometer is fed by fibers optimized for the respective wavelength range. Such a fiber-fed design provides a natural solution for the implementation of the different observing modes discussed in the science case (high resolution, moderate multiplexing at intermediate resolution and IFU+AO). The different modes are obtained by feeding the pseudo-slit in the focal plane of the spectrometers with systems of fibers deployed in different spatial configurations.

Fig.A1 shows the possible scheme of different modules: up to four, totally independent spectrometers, optimized in the UB+V and R+I bands and (cryogenic) in the Y+J+H and K bands, fed by a set of optimized fibers.

We note that the standard telecommunication (low-OH) fibers in the YJH bands ($0.9\mu m < \lambda < 1.8\mu m$) are so transparent that they can be up to 300m long, without affecting the performance of the system. Hence this spectrometer can be accommodated nearly anywhere in the telescope enclosure. The fibers for the blue and K bands, instead, need to be shorter than about 25m to have good throughput, implying that these spectrometers would probably have to sit on one of the Nasmyth platforms. To better clarify the issue associated with the fiber length (hence location of the individual spectrographs), Fig.A2 shows the maximum length of various fibers types, as a function of wavelength, to achieve a transmission higher than 90%.

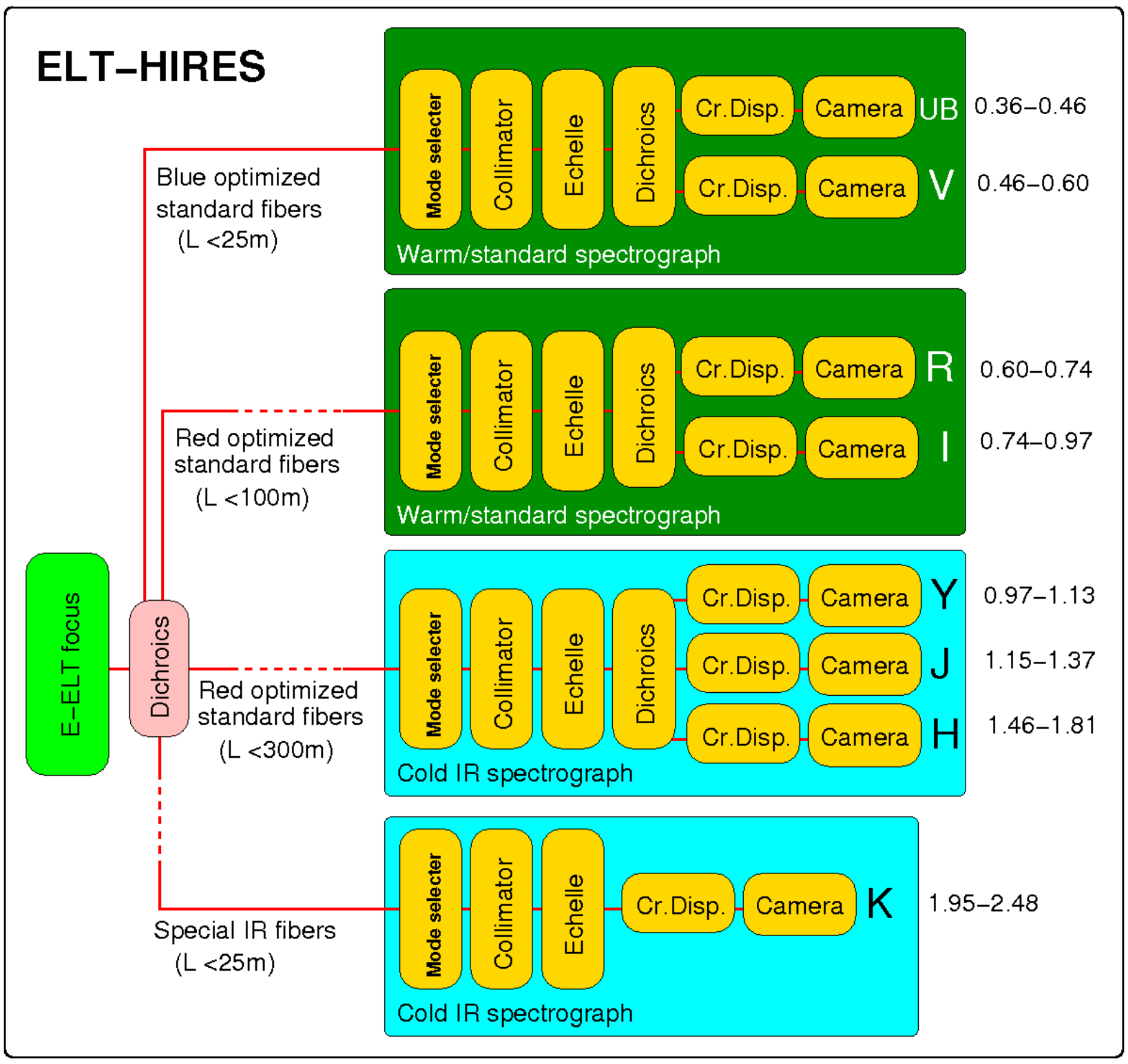

Fig.A1. Possible scheme showing the distribution of the spectroscopic modules for HIRES.

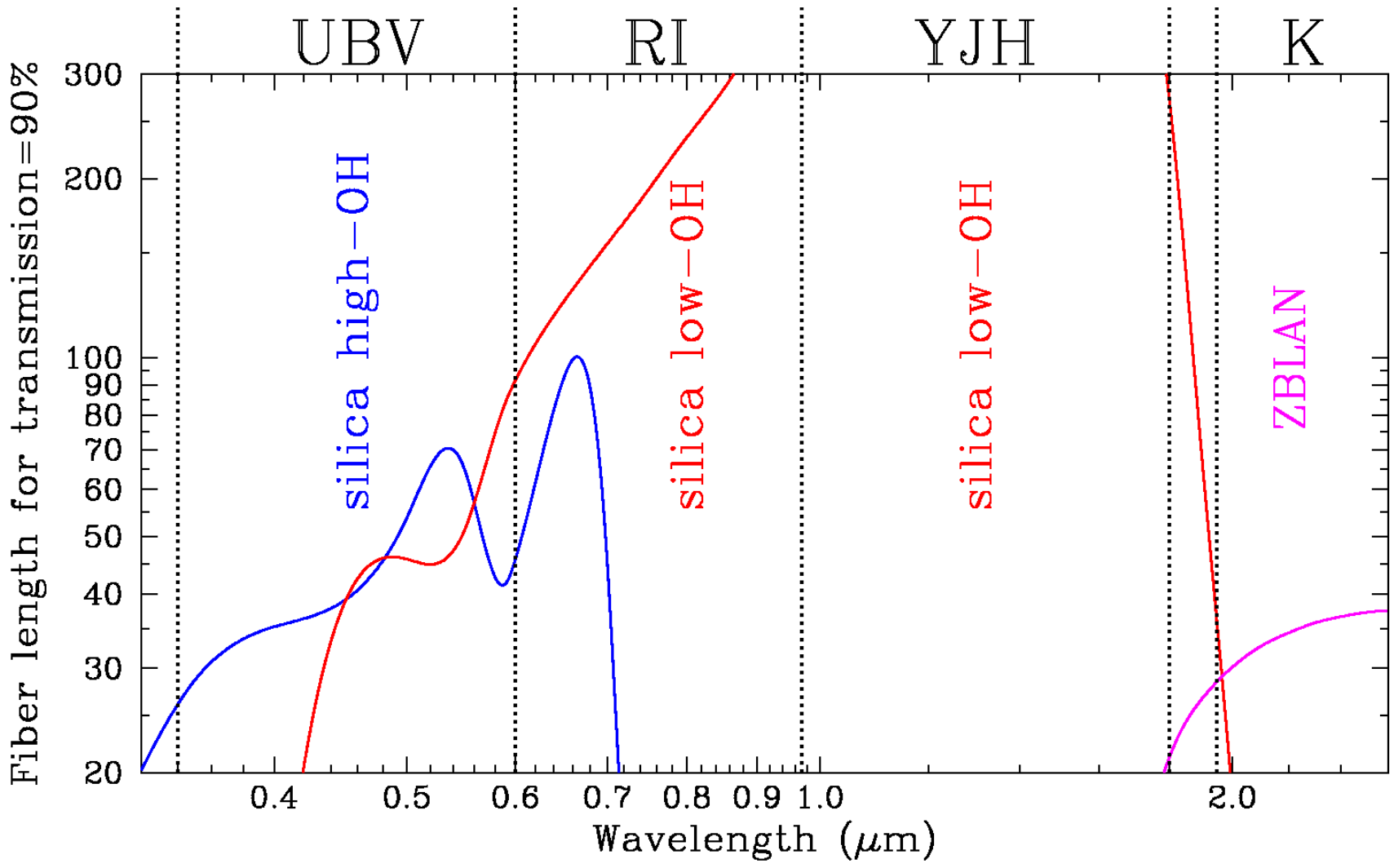

Fig.A2. The curves show, as a function of wavelength, the maximum fiber length to achieve an internal transmission higher than 90% for different type of fibers. This also illustrates the need of different modules (spectrometers) to cover the full wavelength range. Note that for the near-IR non-thermal spectral range (0.9 to 1.8μm) the length of the fibers is not really an issue and the associated spectrometers can be located beyond 300m from the focal station. For the blue and K-band spectral ranges the maximum length of the fibers should be about 25m.

Fig.A3 shows how the different observing modes can be realized leaving the optics and gratings within the spectrometers fixed, by simply changing the pseudo-slit illumination through a different deployment and feeding of the fibers. In the basic "high resolution mode", the light is collected from the telescope by two ~0.8" fibers, one for the object and the other for the sky or calibration (a three fibers system can be envisaged, with two fibers allocated for the sky, so to reduce the sky photon noise). The light from the fibers is sliced into 2x6 rectangular images aligned along the spectrometer pseudo-slit. Only with the slicing technique is it possible to achieve the required spectral resolution (R~100,000) with a relatively "small" grating (~1.2m in size) even if the observations are seeing-limited.

In the medium-resolution mode with moderate multiplex the light is collected from the focal plane of the telescope by 10 fibers (~0.9" in diameter), without any slicing. The fibers images are directly arranged onto the pseudo-slit of the spectrometer. Without changing anything within the spectrometer, this setup automatically delivers a spectral resolution of about 15,000. The ten fibers can be all allocated to targets or, for faint objects, half of them can be allocated to the sky.

A high spectral and spatial resolution IFU mode can be realized with a bundle of fibers fed at the focal plane of a dedicated SCAO module, or at the focus of an already existing AO module (e.g. MAORI). The size of the individual fibers would match the diffraction limit in the preferred near-IR band. Inside the spectrograph, the images of the individual IFU fibers would be deployed along the pseudo-slit of the spectrograph.

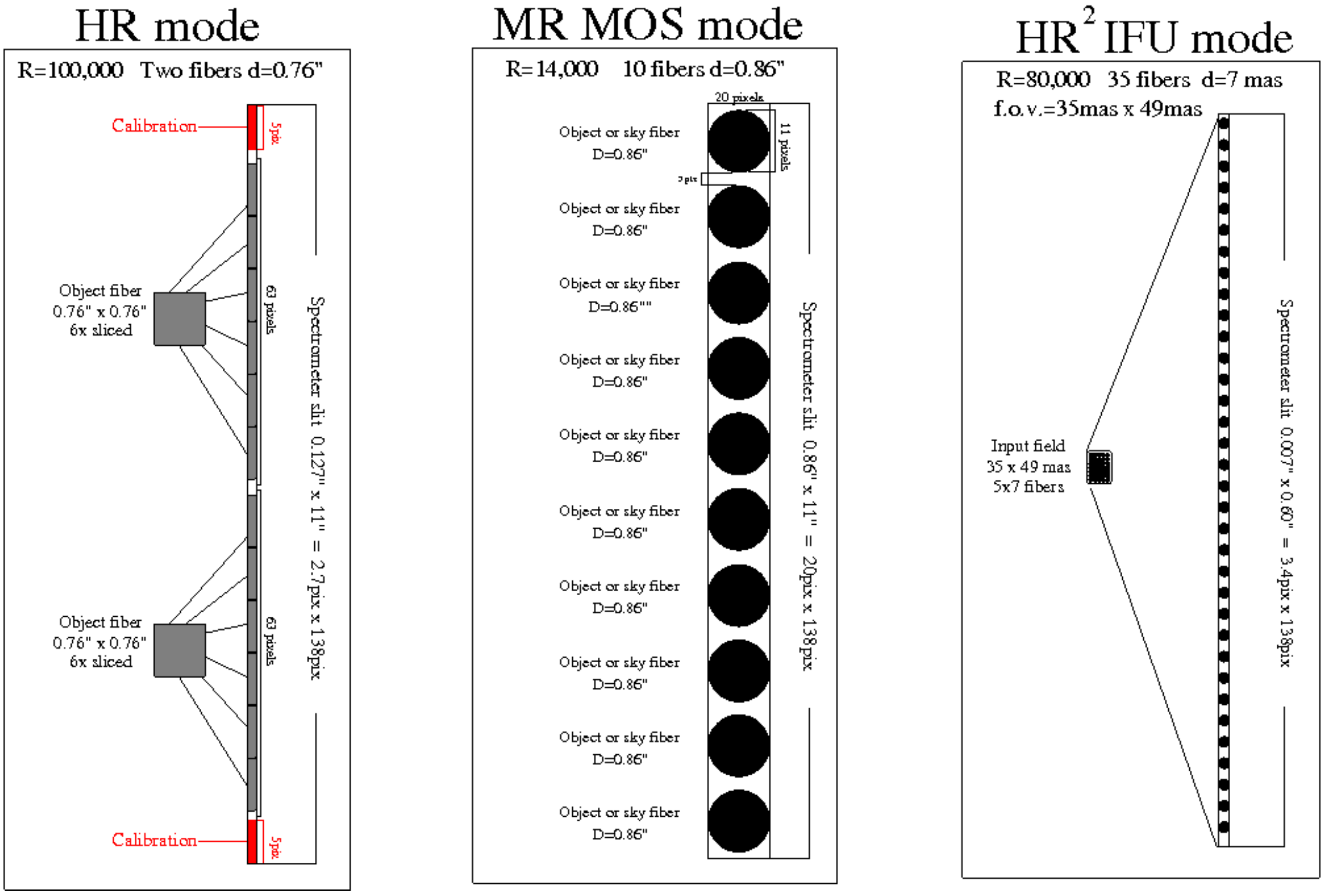

Fig.A3. Illustration of the different possible illuminations of the spectromter pseudo-slit through different set of fibers.

The different observing modes can be selected by a system of two mirrors that relay the light from the different set of fibers to the pseudo-slit of the spectrometer. As illustrated in Fig.A4, the system can be realized in such a way that the basic high-resolution mode is the one feeding directly the spectrometer without any relay-mirror. This guarantees that the stability of this basic mode is not affected by the presence of the other observing modes. The other two modes (medium-resolution with moderate multiplexing and IFU+AO), which do not have stringent stability constraints, are enabled by swapping relay mirrors into the beam.

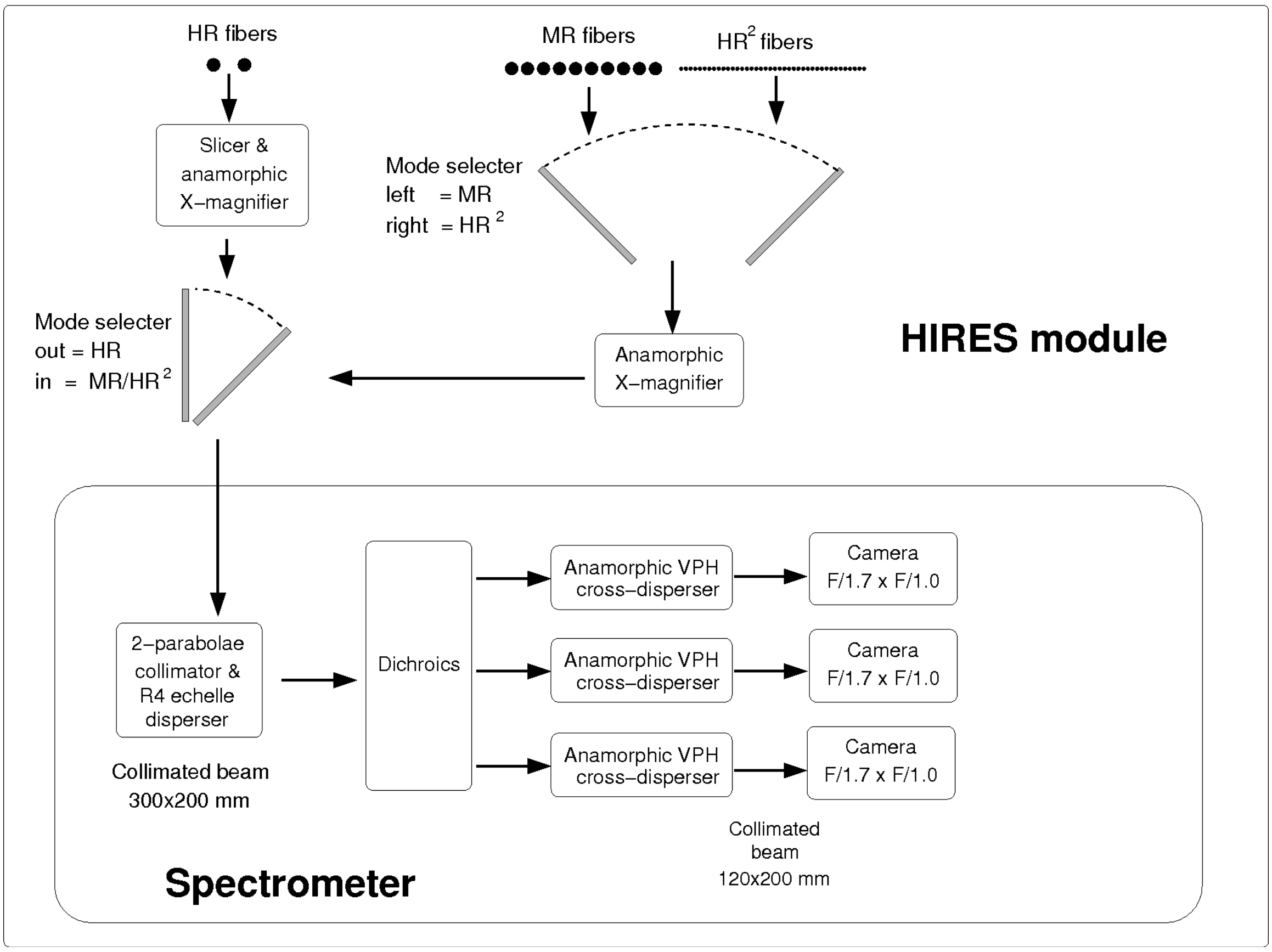

Fig.A4. Block diagram for a possible module of HIRES, also illustrating the mirrors relay system to enable different observing modes.

## *Affiliations:*

[1] Cavendish Laboratory, University of Cambridge, 19 J. J. Thomson Ave., Cambridge CB3 0HE, UK
[2] Kavli Institute for Cosmology, University of Cambridge, Madingley Road, Cambridge CB3 0HA, UK
[3] Institute of Astronomy, Madingley Road, Cambridge, UK, CB3 0HA
[4] Centre for Astrophysics and Supercomputing, Swinburne University of Technology, Hawthorn, Melbourne, Victoria 3122, Australia
[5] INAF-Osservatorio Astronomico di Bologna, Via Ranzani 1, I-40127 Bologna, Italy
[6] INAF - Osservatorio Astronomico di Capodimonte, via Moiariello 16, 80131, Napoli, Italy
[7] Physikalisches Institut, University of Bern, Sidlerstrasse 5, 3012 Bern, Switzerland
[8] Instituto de Astrofísica de Andalucía-CSIC, Glorieta de la Astronomía s/n. P.O. Box 3004 E-18080, Spain
[9] Instituto de Astrofísica de Canarias, 38205, La Laguna, Tenerife, Spain
[10] Thüringer Landessternwarte Tautenburg, Sternwarte 5, 07778, Tautenburg, Germany
[11] Australian National University, Research School of Astronomy and Astrophysics, Cotter Road, Weston ACT 2611, Australia
[12] Department of Physics & Astronomy, University of Leicester, University Road, Leicester LE1 7RH, UK

[13] UJF-Grenoble 1/CNRS-INSU, Institut de Planétologie et d'Astrophysique de Grenoble (IPAG) UMR 5274, 38041, Grenoble, France
[14] Laboratoire d'Astrophysique de Marseille, UMR 6110 CNRS, Universit de Provence, 38 rue Fr´ed´eric Joliot-Curie, 13388 Marseille Cedex 13, France
[15] Institut d'Astronomie et d'Astrophysique, Université Libre de Bruxelles, CP. 226, Boulevard du Triomphe, 1050, Bruxelles, Belgium
[16] Dipartimento di Fisica e Astronomia, Università di Bologna, V.le Berti Pichat 6/2, I-40127 Bologna, Italy
[17] Institute for Astronomy, University of Edinburgh, Royal Observatory, Edinburgh EH9 3HJ
[18] INAF-OATS, Via Tiepolo 11, 34143 Trieste, Italy
[19] Lund Observatory, Box 43, SE-22100 Lund, Sweden
[20] European Southern Observatory, Karl-Schwarzschild-Str. 2, 85748 Garching, Germany
[21] Centro de Astrofísica, Universidade do Porto, Rua das Estrelas, 4150-762, Porto, Portugal
[22] Dark Cosmology Centre, Niels Bohr Institute, Copenhagen University, Juliane Maries Vej 30, Copenhagen O, Denmark
[23] Department of Physics, University of Warwick, Coventry, CV4 7AL, UK
[24] Institut d'Astrophysique et Géophysique, Université de Liège, allée du 6 Août 17, 4000, Liège, Belgium
[25] Department of Physics and Astronomy, Division of Astronomy and Space Physics, Box 515, 751 20 Uppsala, Sweden
[26] Université de Nice Sophia Antipolis, CNRS, Observatoire de la Côte d'Azur, bd. de l'Observatoire, BP 4229, 06304, Nice Cedex 4, France
[27] Department of Astrophysics / IMAPP, Radboud University Nijmegen, PO Box 9010, 6500 GL, Nijmegen, The Netherlands
[28] Laboratoire Lagrange (UMR 7293), Université de Nice Sophia Antipolis, CNRS, Observatoire de la Côte d'Azur, BP 4229, 06304, Nice Cedex 4, France
[29] Observatoire Astronomique de l'Universite de Geneve, 51 Ch. des Maillettes, - Sauverny - CH1290, Versoix, Suisse
[30] Dipartimento di Astronomia e Scienza dello Spazio, Università degli Studi di Firenze, Largo E. Fermi 2, 50125 Firenze, Italy
[31] INAF - Osservatorio Astronomico di Roma, via di Frascati 33, 00040, Monte Porzio Catone, Italy
[32]INAF - Osservatorio Astrofisico di Arcetri, Largo E. Fermi 5, 50125, Firenze, Italy
[33] Université Paris 6, Institut d'Astrophysique de Paris, CNRS UMR7095, 98bis bd Arago, 75014 Paris, France
[34] Universität Göttingen, Institut für Astrophysik, Friedrich-Hund-Platz 1, 37077, Göttingen, Germany
[35] Departamento de Física Estelar, Instituto de Astrofísica de Andalucía (IAA-CSIC), 18008 Granada, Spain
[36] Max Planck Institute for Extraterrestrial Physics, 85748 Garching bei München, Germany
[37] Leiden Observatory, Leiden University, Postbus 9513, 2300-RA Leiden, The Netherlands
[38] Leibniz-Institute for Astrophysics Potsdam (AIP) An der Sternwarte 16, D-14482 Potsdam, Germany
[39] Department of Physics, University of Warwick, Coventry CV4 7AL, UK
[40] Kapteyn Astronomical Institute, University of Groningen, Postbus 800, 9700AV, Groningen, the Netherlands
[41] Department of Electrical Engineering and Center of Astro Engineering, Pontificia Universidad Catolica de Chile, Av. Vicuña Mackenna 4860 Santiago, Chile